\documentclass[preprintnumbers,superscriptaddress,showkeys,showpacs,byrevtex]{revtex4}
\usepackage{amsmath,amsfonts,amssymb,amscd,amsxtra,amsthm}
\usepackage{graphicx}
\usepackage{bm}
\usepackage{multirow}
\newcommand{\nn}{\nonumber}

\newcommand {\be} {\begin{equation}}
\newcommand {\ba} {\begin{eqnarray}}
\newcommand {\ee} {\end{equation}}
\newcommand {\ea} {\end{eqnarray}}
\begin{document}
\preprint{CYCU-HEP-19-13}
\title{Empirical estimate of the unpolarized dihadron fragmentation functions }
\author{Dong-Jing Yang}

\email[E-mail: ]{djyang@std.ntnu.edu.tw}
\affiliation{Department of Physics, National Taiwan Normal University,
Taipei 10610, Taiwan}
\author{Chung Wen Kao}
\email[E-mail  (Corresponding Author): ]{cwkao@cycu.edu.tw}
\affiliation{Department of Physics and Center for High Energy Physics, Chung-Yuan Christian University, Chung-Li 32023, Taiwan}
\author{Seung-il Nam}
\email[E-mail: ]{sinam@pknu.ac.kr}
\affiliation{Department of Physics and Institute for Radiation Science \& Technology (IRST), Pukyong National University (PKNU), Busan 608-737, Republic of Korea}
\affiliation{Center for Extreme Nuclear Matters (CENuM), Korea University, \\Seoul 02841, Republic of Korea}
\affiliation{Asia Pacific Center for Theoretical Physics (APCTP), Pohang 790-784,
Republic of Korea}
\date{\today}
\begin{abstract}
We estimate the unpolarized dihadron fragmentation functions (uDiFFs) within the framework of the single-cascade jet algorithm.
We first obtain the elementary fragmentation functions generating the single hadron fragmentation functions by using
the single-cascade jet algorithm.
These results are very close to the empirical parametrizations of the single hadron fragmentation functions which are extracted from the experimental data of the $e^+e^-$ annihiliation
and semi-inclusive deeply inelastic scatterings (SIDIS).
We then use those elementary fragmentation functions to generate the dihadron fragmentation functions by the help of the single-cascade jet algorithm again.
The comparison between our empirical results with the results of the Nambu-Jona-Lasinio model and the non-local chiral quark model (NL$\chi$QM) is also presented.

\end{abstract}
\pacs{12.38.Aw, 13.60.-r, 12.39.-x, 14.40.Aq, 11.10.Hi.}
\keywords{dihadron fragmentation functions, Semi-inclusive DIS, empirical parametrizations. }
\maketitle
\section{Introduction}
The single-hadron fragmentation functions (SiFFs) play very important roles in the analysis of the scattering processes involving hadrons,
such as the Collins fragmentation function, which describes the hadronization of a transversely polarized quark.
Therefore, SiFFs have been intensively studied ~\cite{Book}.\\

Among various kinds of SiFFs, the most basic one is
the unpolarized single-hadron fragmentation function $D^{h}_{q}(z,Q^2)$ (uSiFF). It is defined
to characterize the process of an unpolarized quark $q$ with the virtuality $Q^2$ to hadronize into a hadron $h$
carrying a fraction of light-cone momentum $z$. In principle, the uSiFF can be extracted from experimental data of
semi-inclusive processes such as $e^{+}+e^{-}\to h+X$ or $e^-+p\to e+h+X$ provided that some extra assumptions,e.g., only the leading twist contribution is taken into account.

One natural extension of uSiff is the unpolarized dihadron fragmentation functions (uDiFF).
When analyzing the semi-inclusive processes with two specified hadrons in the final states such as $e^{+}+e^{-}\to h_{1}+h_{2}+X$ or $e^-+p\to e+h_{1}+h_{2}+X$,
one needs to define the unpolarized dihadron fragmentation function $D^{h_1,h_2}_{q}(z_1, z_2, Q^2)$ (uDiFF) which is the probability of a quark $q$ fragmenting into
two hadrons $h_1$ and $h_2$ with the light-cone momentum fractions $z_1$ and $z_2$, respectively~\cite{Konishi78}.
The QCD evolution equations of uDiFFs have also been intensively investigated~\cite{Vendramin81,Sukhatme80,deFlorian04,Majumder04,
Majumder05}.\\

\begin{figure}[t]
\begin{tabular}{c}
\includegraphics[width=8.5cm]{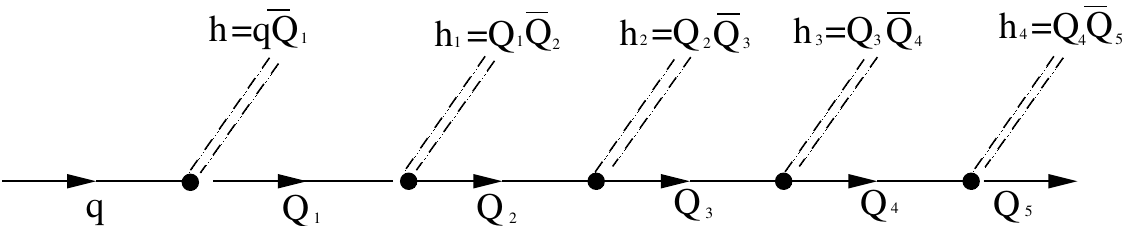}
\end{tabular}
\caption{Quark fragmentation cascade process.}
\label{cascade}
\end{figure}

So far there is no empirical extraction of uDiFFs from experimental data. Only some model calculations are available.
Those model calculations all base on the single-cascade jet algorithm presented in Fig.~(\ref{cascade}). The models provide the single-step elementary fragmentation functions
$d_q^h(z)$ describing the emission of a single hadron, $q\rightarrow h(q\bar{Q})+Q$, here the quark content of the emitted hadron is $(q\bar{Q})$.
Usually, emitted hadrons are assumed to be pseudoscalar mesons only, it is just for simplicity.
One then solve the following coupled-channel integral equations to obtain the physical single hadron fragmentation functions $D^{h}_{q}(z)$:
\begin{equation}
D^{h}_{q}(z)dz=\hat{d}^{h}_{q}(z)dz+\sum_{Q}\int^{1}_{z}dy\,
\hat{d}^{Q}_{q}(y)D^{h}_{Q}\left(\frac{z}{y}\right)\frac{dz}{y}.
\label{multijet}
\end{equation}
Here $D^{h}_{q}(z)dz$ is the probability for a quark $q$ to emit a hadron $h$
which carries the light-cone momentum fraction from $z$ to $z+dz$.
$\hat{d}^{Q}_{q}(y)dy$ is the probability for a quark $q$ to emit a hadron with flavour composition $q\bar{Q}$ at one step and the final quark becomes
$Q$ with the light-cone momentum fraction from $y$ to $y+dy$. Eq.~(\ref{multijet}) actually describes a fragmentation cascade process of hadron
emissions of a single quark depicted in Fig.~(\ref{cascade}).
With the elementary fragmentation functions $d^h_q(z)$ and the derived $D^h_q(z)$ in hand, one is able to obtain the unpolarized dihardon
fragmentation functions by solving the following equations,
\begin{equation}
D^{h_{1},h_{2}}_{q}(z_{1},z_{2})=\delta_{aq}\hat{d}^{h_{1}}_{q}(z_{1})\frac{D^{h_{2}}_{q_{1}}\left(\frac{z_{2}}{1-z_{1}}\right)}{1-z_{1}}+
\delta_{bq}\hat{d}^{h_{2}}_{q}(z_{2})\frac{D^{h_{1}}_{q_{2}}\left(\frac{z_{1}}{1-z_{2}}\right)}{1-z_{2}}
+\sum_{Q}\int^{1}_{z_{1}+z_{2}}\frac{d\eta}{\eta^2}\hat{d}^{Q}_{q}(\eta)D^{h_{1},h_{2}}_{Q}\left(\frac{z_{1}}{\eta},\frac{z_2}{\eta}\right).
\label{Dihadron1}
\end{equation}

In other words, once the $d^h_q(z)$ are determined, $D^h_q(z)$ and $D^{h_1,h_2}_q(z_1,z_2)$ are also determined by Eq.(\ref{multijet})
and Eq.~(\ref{Dihadron1}).
To make an excellent estimate of the uDiFF $D^{h_1,h_2}_q(z_1,z_2)$, it is necessary to find the adequate $d^h_q(z)$
to produce $D^h_q(z)$ which agree well with the ones given by the empirical parametrizations. Those empirical
parametrizations are obtained by the global fit of many experimental data.\\

Naively one may hope that the results of $D^h_q(z)$ from previous work based on the models would shed some light on the forms of
$d^h_q(z)$. Unfortunately, such an expectation is
frustrated by the diversity of the model results.
For example, the NJL model provide a much better description
of the kaon fragmentation functions in general. On the other hand, the NL$\chi$QM provide superior results of the pion fragmentation functions. It is also impossible to take different elementary fragmentation functions from different models
because that Eq.(\ref{multijet}) are couple-channel equations.\\

Since the models are not the reliable sources of the elementary fragmentation functions, one needs to find the elementary fragmentation functions by brute force. Our goal is to invert the
Eq.(\ref{multijet}) to find the $d^h_q(z)$ by using the empirical
parametrizations of the $D^h_q(z)$ as the inputs.
It turns out to be impossible to achieve this goal for all channels. Therefore, we are satisfied ourself to find the elementary fragmentation functions which can reproduce the
SiFFs of the favoured channels in a reasonably well agreement with the empirical parametrization, and apply them in the single-cascade jet algorithm to generate the uDiFFs.\\

Furthermore we compare our empirical results of the uDiFFs with the results of the model calculations. We notice that our empirical results are significantly different from the model calculations and consistent with the conclusion of our discovery that the NJL model underestimates the pion fragmentation functions and the NL$\chi$QM underestimate the
kaon fragmentation functions.\\

\begin{figure}[b]
\begin{tabular}{cc}
\includegraphics[width=6.2cm]{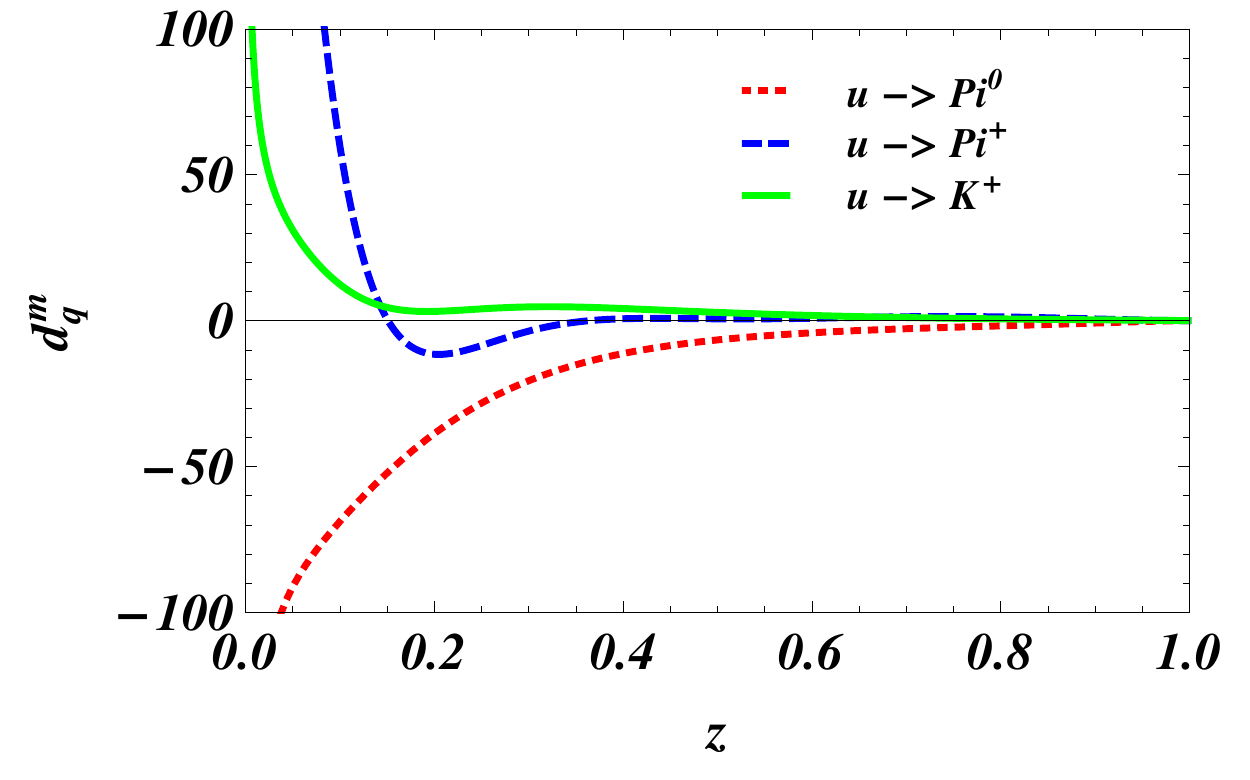}
\includegraphics[width=6.2cm]{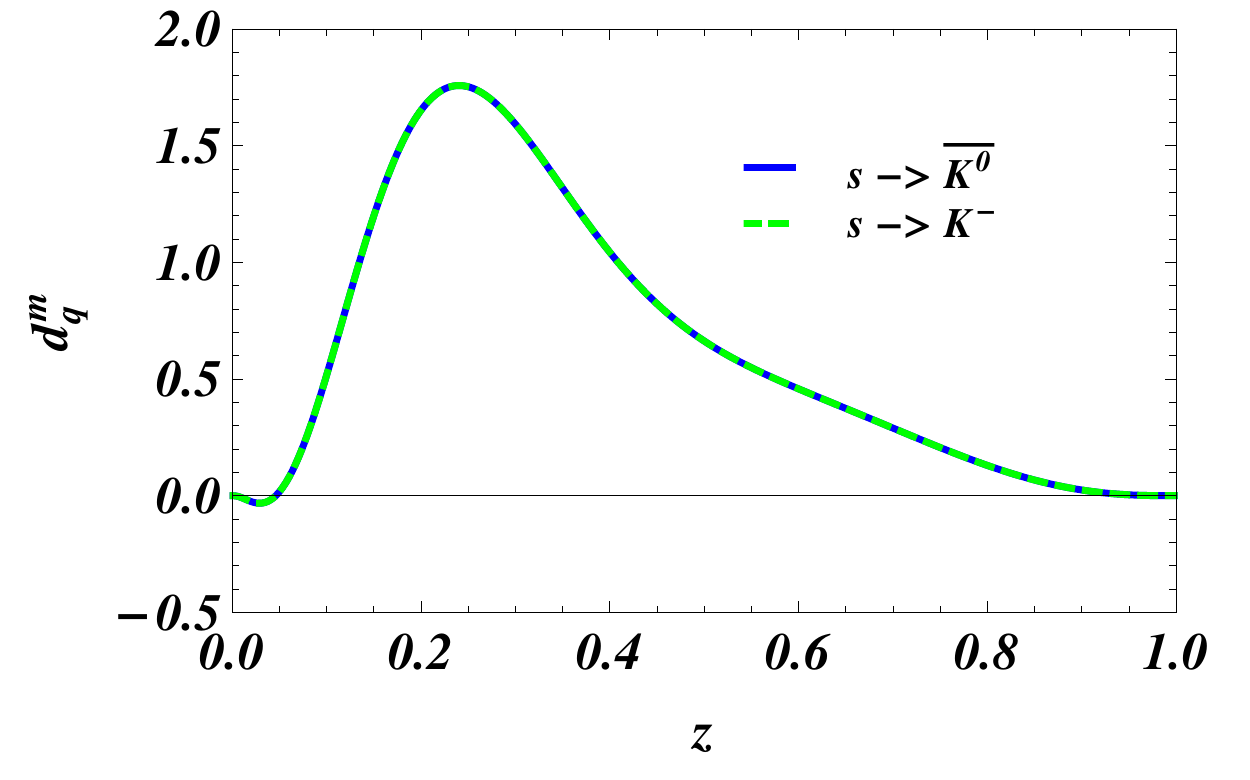}
\end{tabular}
\caption{The result of inverted result of the elementary fragmentation functions with the $D(z)$ without modification as the input.}
\label{invert}
\end{figure}
This article is organized as follows: We introduce the process of finding the elementary fragmentation functions
which are able to reproduce the favoured channels uSiFFs close to the empirical parametrization in Sec.~II.
In Sec.~III we present and discuss our results of uDiFFs which have been evolved to $Q^2=4\,\mathrm{GeV}^2$ .
Finally, we present our conclusion and outlooks in Sec.~IV.\\
\begin{figure}[t]
\begin{tabular}{cc}
\includegraphics[width=5.2cm]{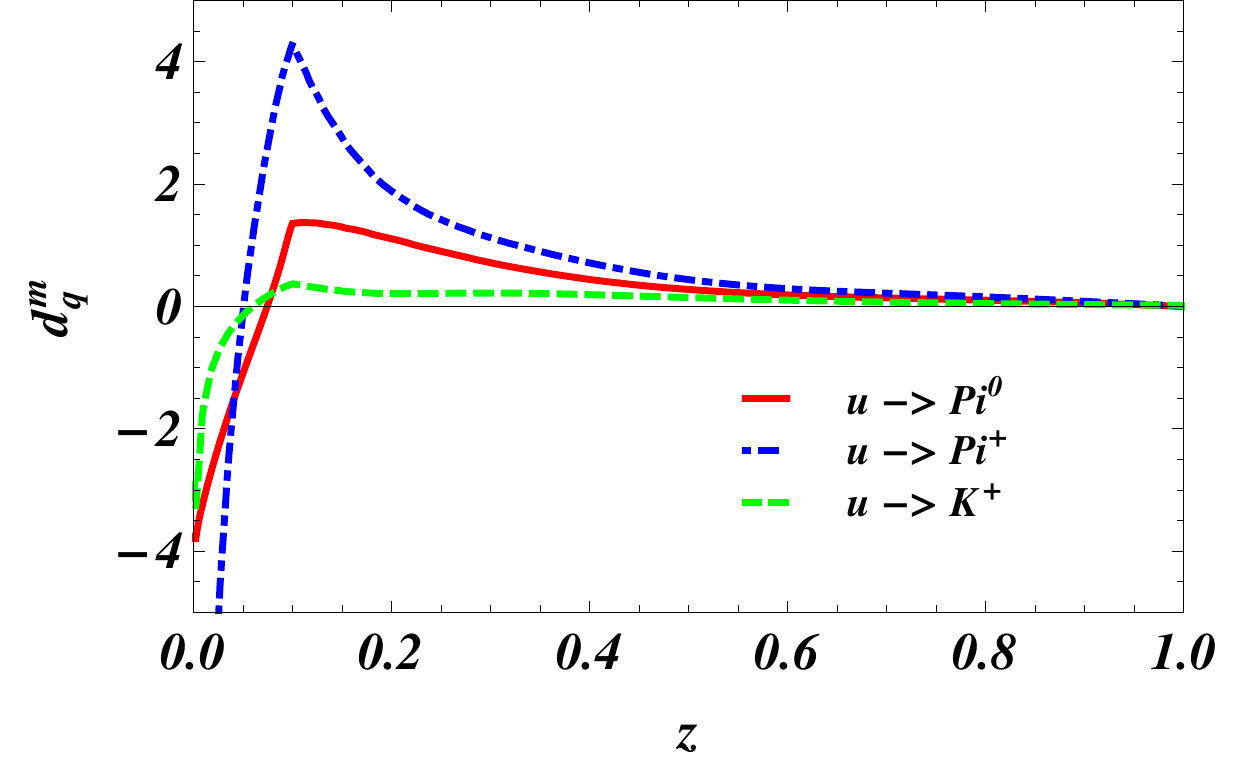}
\includegraphics[width=5.2cm]{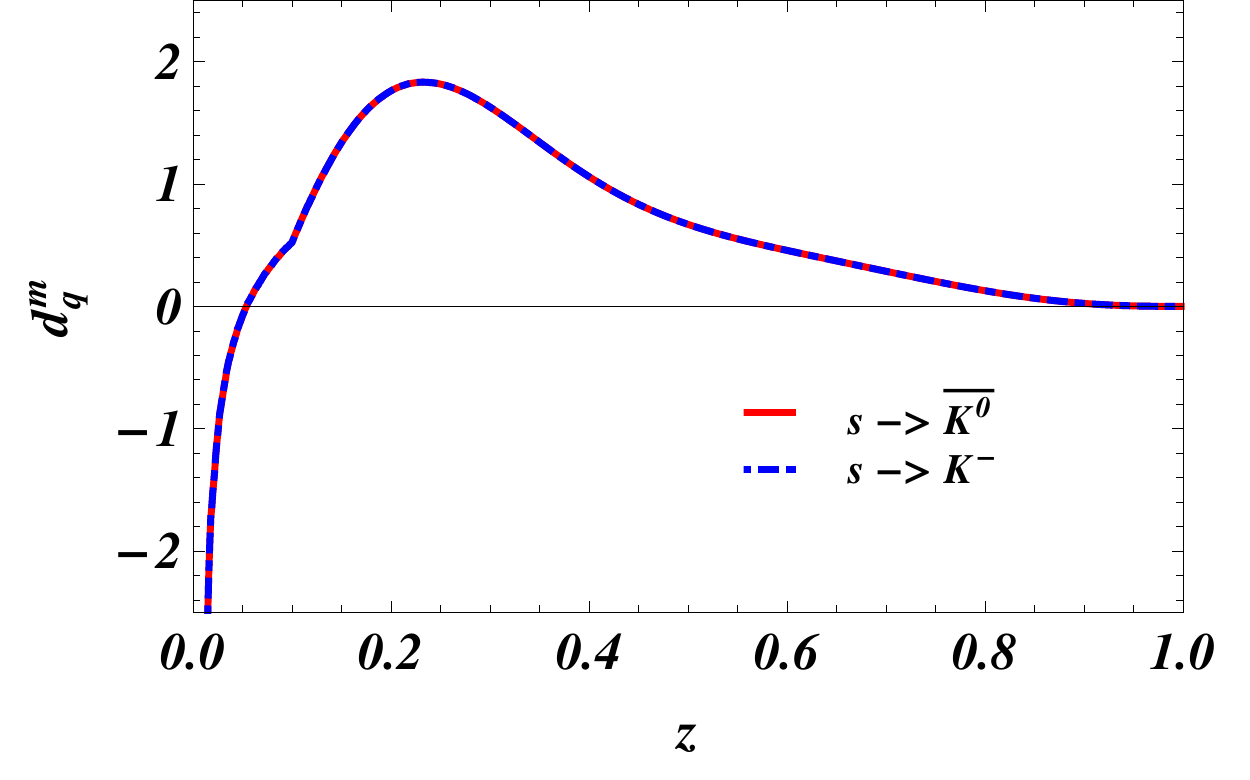}
\end{tabular}
\caption{The result of inverted result of the elementary fragmentation functions with the $D(z)$ with the modification as the input.}
\label{curve}
\end{figure}

\section{Elementary fragmentation functions in the single cascade algorithm }
Our first step is to find the elementary fragmentation functions $d^{h}_{q}(z)$ to reproduce the single-hadron fragmentation functions
$D^{h}_{q}(z)$ provided by the empirical parametrizations. Within the algorithm of a single-cascade jet depicted in Fig.~(\ref{cascade}),
one can connect $d_q^h(z)$ and $D_q^h(z)$ by solving Eq.~(\ref{multijet}).\\

In this article we assume that the fragmentation functions own the exact isospin symmetry .
Consequently, there are the following relations among these fragmentation functions,
\begin{eqnarray}
d^{\pi^{+}}_{u}(z)&=&d^{\pi^{-}}_{d}(z),\,\,d^{\pi^{0}}_{u}(z)=d^{\pi^{0}}_{d}(z),\,\,d^{K^{+}}_{u}(z)=d^{K^{0}}_{d}(z),d^{K^{-}}_{s}(z)=d^{\bar{K^{0}}}_{s}(z),\nn \\
D^{\pi^{+}}_{u}(z)&=&D^{\pi^{-}}_{d}(z),\,\,D^{\pi^{0}}_{u}(z)=D^{\pi^{0}}_{d}(z),\,\,D^{K^{+}}_{u}(z)=D^{K^{0}}_{d}(z),D^{\bar{K}^{0}}_{s}(z)=D^{K^{-}}_{s}(z).\nn
\end{eqnarray}
Note that the above quantities are called "favoured" since the emitted hadron contains the parent quark.
The unfavoured fragmentation functions derived from Eq.(\ref{multijet}) naturally also hold the isospin symmetry:
\begin{eqnarray}
D^{\pi^{-}}_{u}(z)&=&D^{\pi^{+}}_{d}(z),\,\,D^{K^{0}}_{u}(z)=D^{K^{+}}_{d}(z),\,\,D^{\bar{K}^{0}}_{u}(z)=D^{K^{-}}_{d}(z),\nn \\
D^{\bar{K}^{0}}_{d}(z)&=&D^{K^{-}}_{u}(z),\,\,D^{\pi^{+}}_{s}(z)=D^{\pi^{-}}_{s}(z),\,\,D^{K^{+}}_{s}(z)=D^{K^{0}}_{s}(z).\nn
\end{eqnarray}

In this work the emitted hadrons are only limited to the pseudoscalar mesons, therefore,
we have four independent $d^h_q(z)$, which are $d^{\pi^{+}}_{u}(z),d^{\pi^{0}}_{u}(z),d^{K^{+}}_{u}(z)$ and $d^{K^{-}}_{s}(z)$.
We also have four distinct favoured uDiFFs $D^h_q(z)$, they are $D^{\pi^{+}}_{u}(z),D^{\pi^{0}}_{u}(z),D^{K^{+}}_{u}(z)$ and $D^{K^{-}}_{s}(z)$.
Moreover we have seven unfavoured $D^h_q(z)$, they are $D^{\pi^{-}}_{u}(z),D^{K^{0}}_{u}(z),D^{\bar{K^{0}}}_{u}(z),D^{K^{-}}_{u}(z),
D^{\pi^{+}}_{s}(z),D^{K^{+}}_{s}(z)$ and $D^{\pi^0}_{s}(z)$.\\
\begin{figure}[b]
\begin{tabular}{ccc}
\includegraphics[width=5.2cm]{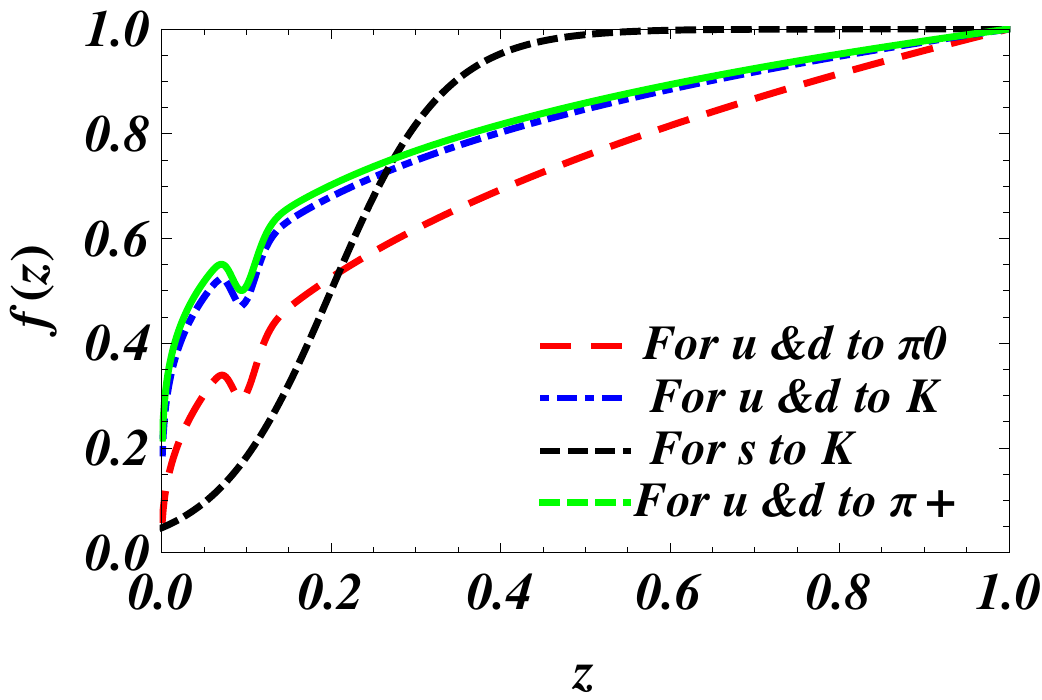}
\includegraphics[width=5.2cm]{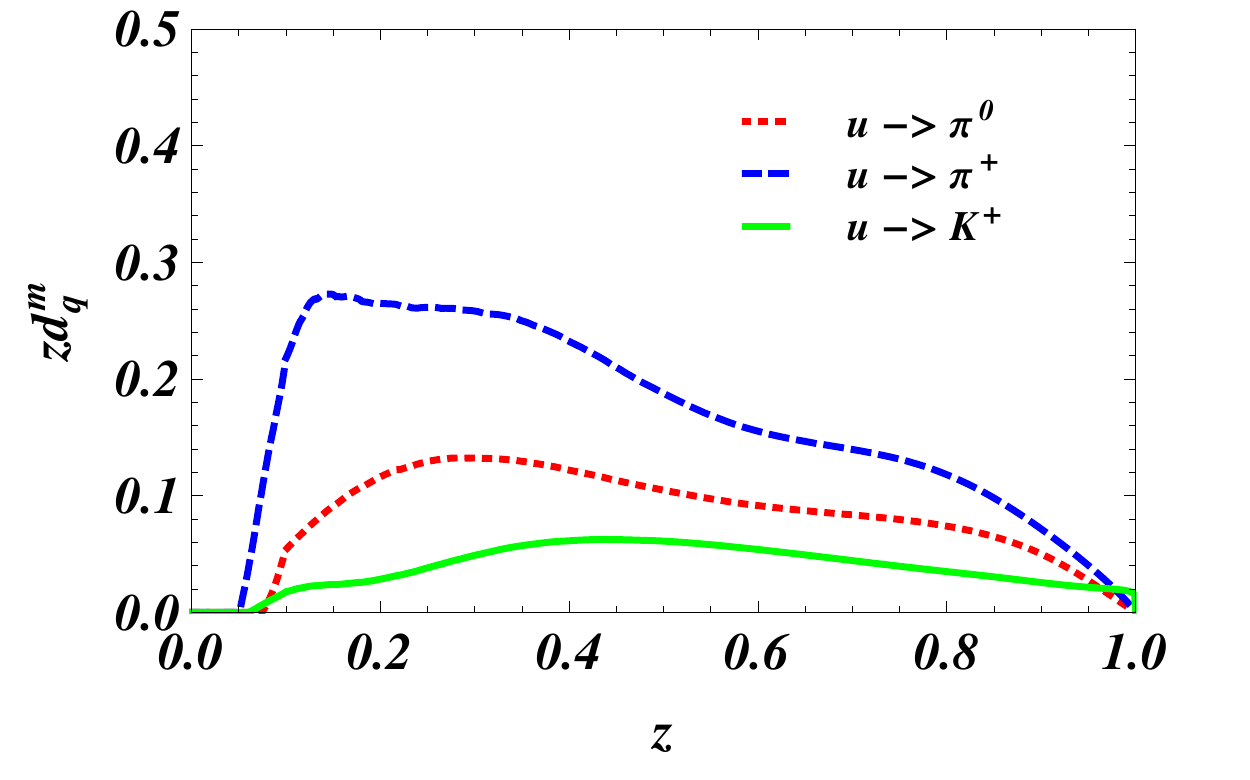}
\includegraphics[width=5.2cm]{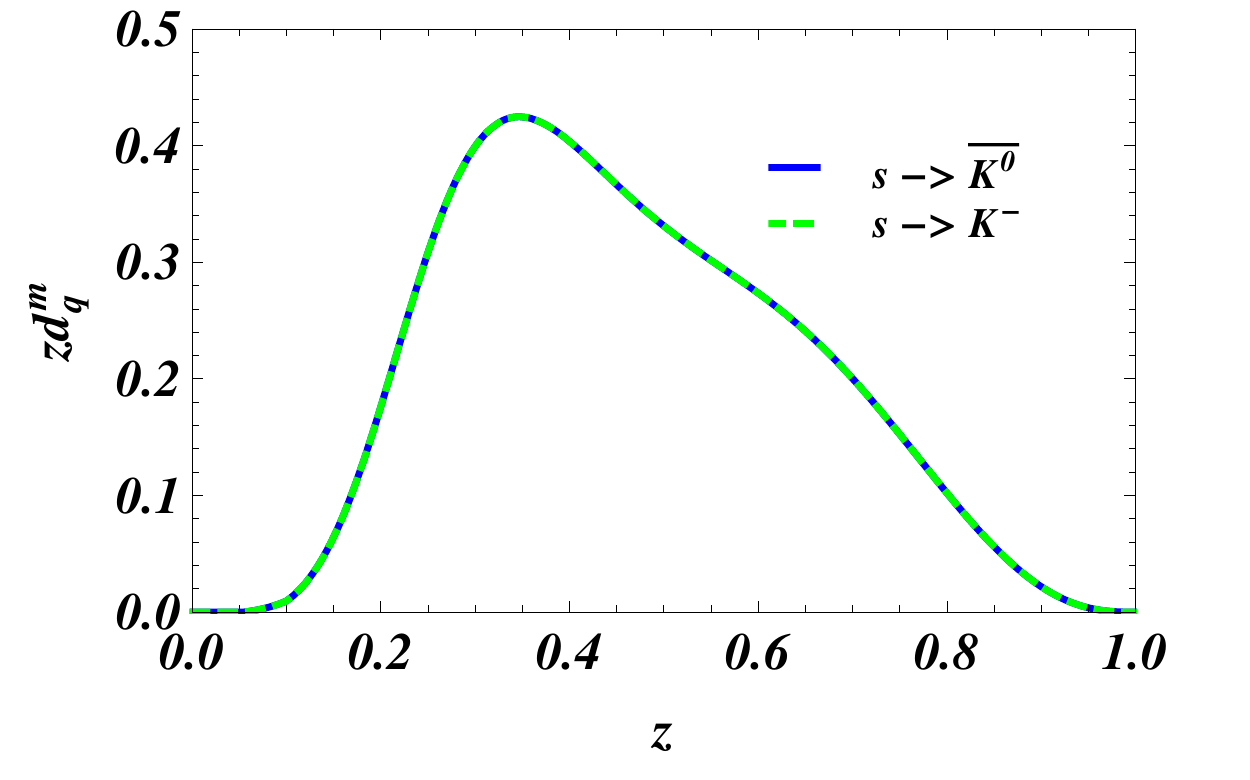}
\end{tabular}
\caption{The result of the elementary fragmentation functions modified by $f(z)$.}
\label{zd}
\end{figure}
To find the elementary fragmentation functions $d^{h}_{q}(z)$ to reproduce the single-hadron fragmentation functions
$D^{h}_{q}(z)$ provided by the empirical parametrizations, the most straightforward way is to invert Eq.~(\ref{multijet}).
Namely we can try to solve the following equations,
\begin{equation}
\hat{d}_q^h(z)=D_q^h(z)-\sum_{h'=q\bar{Q}}{\int_z^1\frac{dy}{y}D_Q^h(\frac{z}{y})\hat{d}_q^{h'}(1-y)}.
\end{equation}
change the variable $(1-y) \rightarrow y$, one obtains,
\begin{equation}
\hat{d}_q^h(z)=D_q^h(z)-\sum_{h'=q\bar{Q}}{\int_0^{1-z}\frac{dy}{1-y}D_Q^m(\frac{z}{1-y})\hat{d}_q^{h'}(y)}.
\label{dD}
\end{equation}
If the values of $D_{q}^{h}(z)$ are taken from the certain empirical parametrizations, then one can
solve Eq.~(\ref{dD}) of the favoured channel.
In this article we choose DSS17 parametrization which is extracted from SIDIS and $e^{+}e^{-}$ annihilation experimental data.
We obtain the results presented in Fig.(\ref{invert}).
The result in Fig.(\ref{invert})
is unlikely to be physical since
they become negative in some regions.
It is due to the fact the $D_q^h(z)$ in the empirical parametrizations is to be assumed to drop to zero when $z$ approaches to zero,
therefore the resultant $d_q^m(z)$ need to be turn to negative in the very low $z$ regime. Hence we modify our input $D_q^m(z)$ by freezing the values of $D_q^m(z<0.1)$ to be $D_q^m(z=0.1)$. It is expected to improve our result.
Such a manipulation is indeed helpful to obtain more reasonable result.\\

There is another
more serious difficulty to overcome to reach our goal.
As a matter of fact, Eq.(\ref{dD})are over determined
because there are eleven equations but only four non-zero $d^h_q(z)$.
Four equations are for the favoured channels and another seven equations
are for the unfavoured ones. The equations for the unfavoured channels
in Eq.(\ref{dD}) are
\begin{equation}
0=D_q^h(z)-\sum_{h'=q\bar{Q}}{\int_0^{1-z}\frac{dy}{1-y}D_Q^h(\frac{z}{1-y})\hat{d}_q^{h'}(y)}.
\label{dDunf}
\end{equation}
Since the elementary fragmentation functions are defined to describe the one-step fragmentation process conserving the flavour quantum numbers.
The equations in Eq.(\ref{dDunf}) are actually very restrict constraints on the input $D^m_q(z)$.
Actually no empirical parametrizations of the SiFFs would satisfy these constraints.
To circumvent this obstacle, we have to only look for the elementary fragmentation functions reproducing
the SiFFs of the favoured channels provided the empirical parametrization.
On the contrary, the unfavoured channel ones have to be determined by Eq.~\ref{dDunf}.
Our strategy is as follows, we will first solve the equations of the favoured channels to obtain four
$d_q^h(z)$ with the $D^h_q(z)$ given by the empirical parametrization as the input.
Then we use the resulting $d_q^h(z)$ and the favoured $D^h_q(z)$ of the empirical parametrization as the input to
solve the remain seven equations to obtain the seven unfavoured $D^h_q(z)$.
With the new obtained unfavoured $D^h_q(z)$ and the empirical favoured $D^h_q(z)$ as the inputs one iterates the whole process till the results
of $d^h_q(z)$ and the unfavoured $D^h_q(z)$ become convergent.
Consequently the resultant $d^h_q(z)$ can generate the $D^h_q(z)$ of the favoured channels coinciding with the empirical parametrizations.
But the $D^h_q$ of the unfavoured channels generated form our $d^m_q(z)$ will be different from the empirical parametrizations.
Because that the empirical parametrizations for the unfavoured channels own much lager uncertainties than the ones of the favoured channels,
hence it is reasonable to choose to reproduce the empirical parametrizations of the favoured channels rather than the unfavored ones.\\

\begin{figure}
\begin{tabular}{cc}
\includegraphics[width=5.2cm]{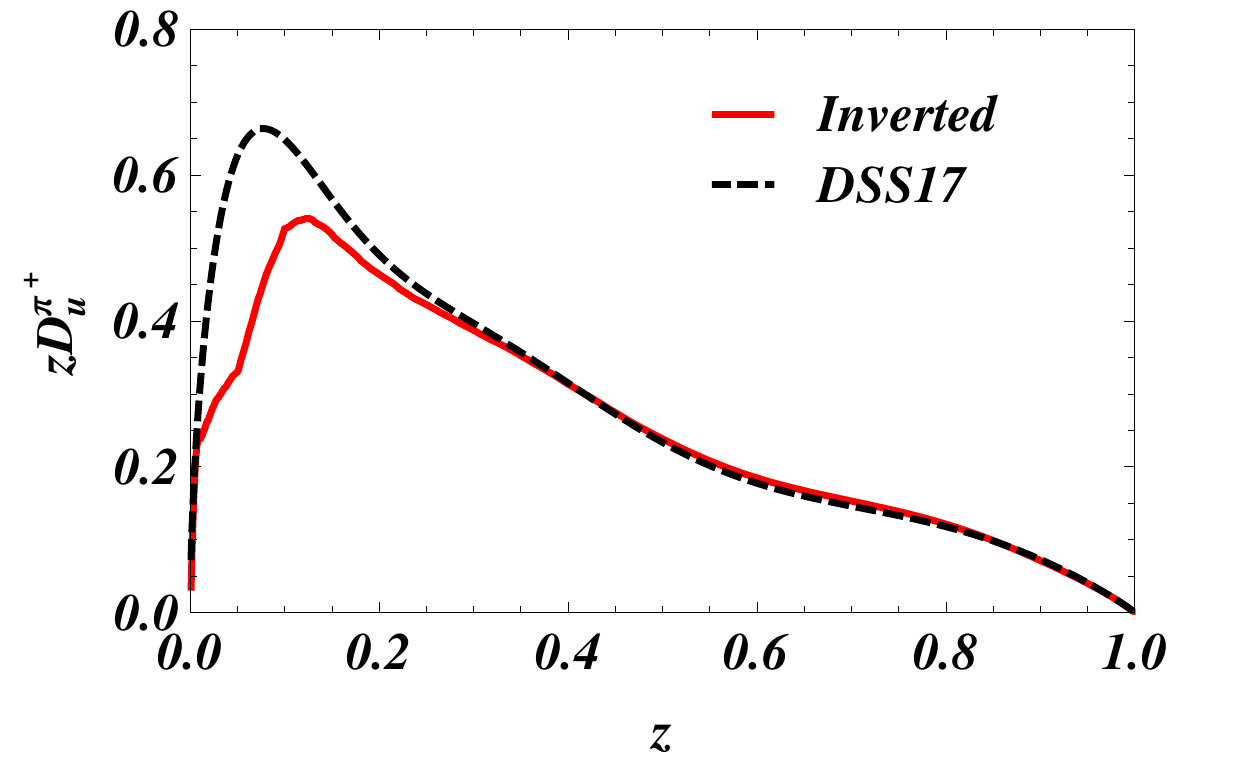}
\includegraphics[width=5.2cm]{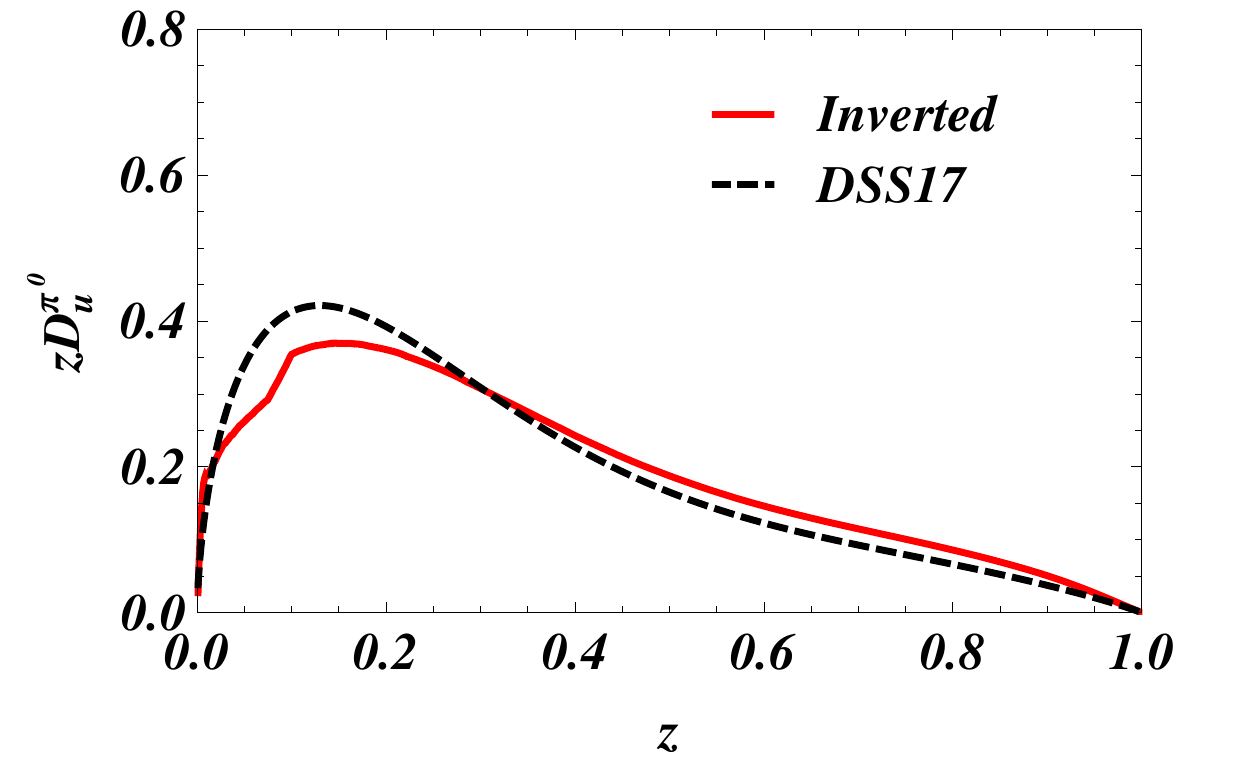}
\end{tabular}
\begin{tabular}{cc}
\includegraphics[width=5.2cm]{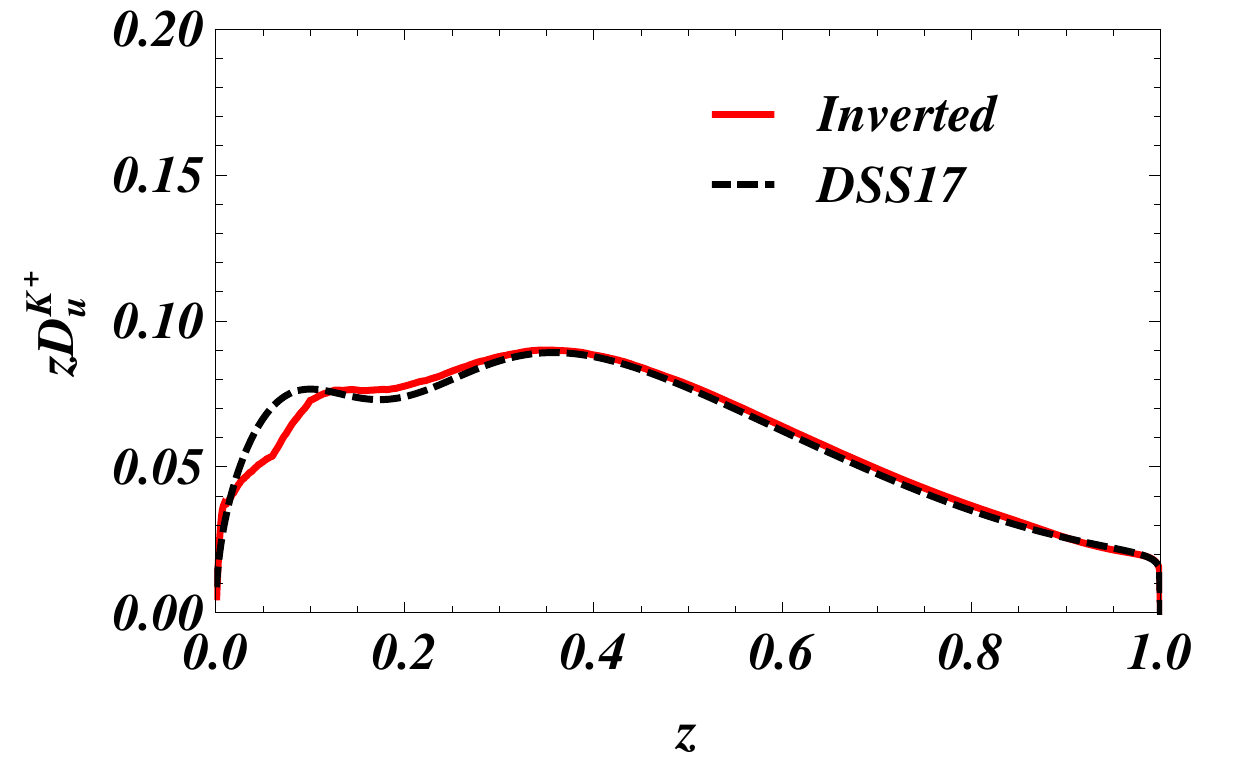}
\includegraphics[width=5.2cm]{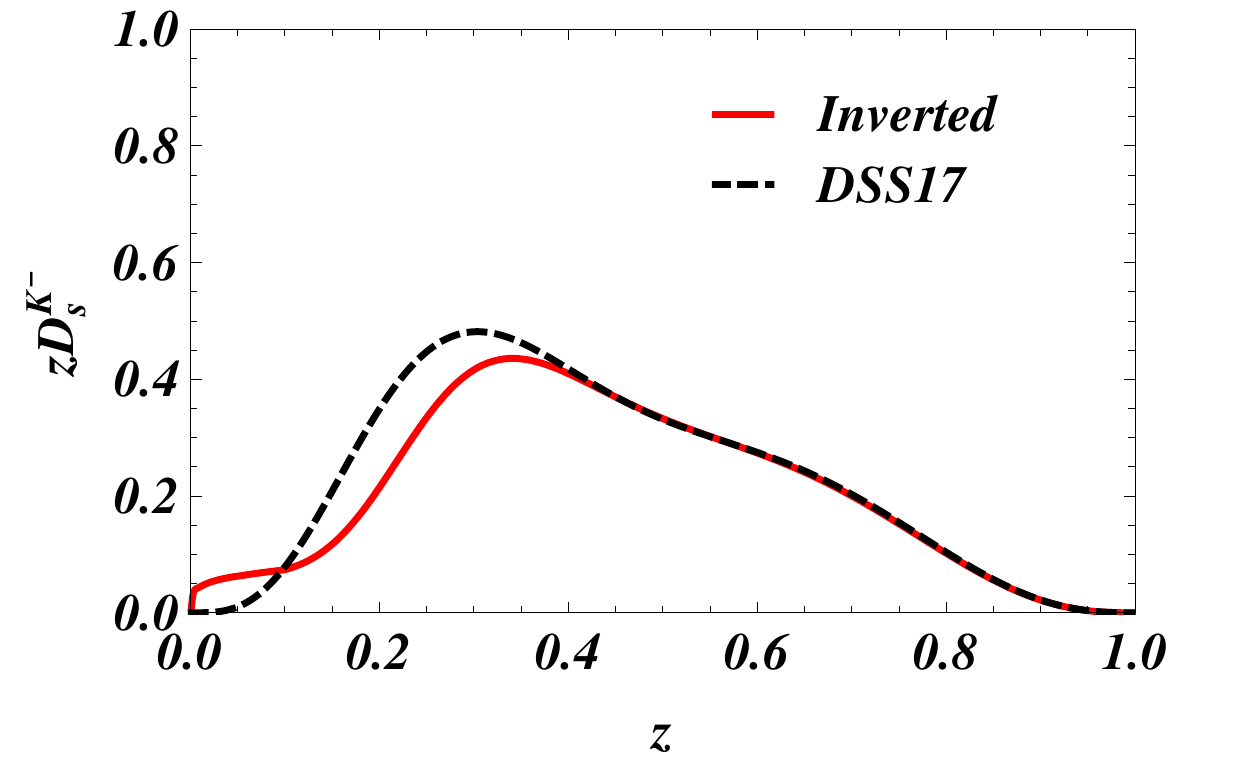}
\end{tabular}
\caption{$z_2 D^{h_1,h_2}_{q}(z_1,z_2)$ with $z_{1}=0.5$ and $Q^2=4\,\mathrm{GeV}^2$ for
(1) $(q,h_1,h_2)=(u,\pi^{+},\pi^{-})$ (left of the top row), (2) $(d,\pi^{+},\pi^{-})$
(middle of the top row), (3) $(s,\pi^{+},\pi^{-})$ (right of the top row),
(4) $(u,\pi^{+},K^{-})$ (left of the middle row), (5) $(d,\pi^{+},K^{-})$
(middle of the middle row), (6) $(s,\pi^{+},K^{-})$ (right of the middle row)
(7) $(u,K^{+},K^{-})$ (left of the bottom row), (8) $(d,K^{+},K^{-})$
(middle of the bottom row), (9) $(s,K^{+},K^{-})$ (right of the bottom row).
The dashed and solid lines denote the results of the NJL-jet model and the nonlocal chiral quark model respectively. The range of $z_2$ is from zero to 0.5.
}
\label{FaSiFF}
\end{figure}
\begin{figure}[t]
\begin{tabular}{cc}
\includegraphics[width=5.2cm]{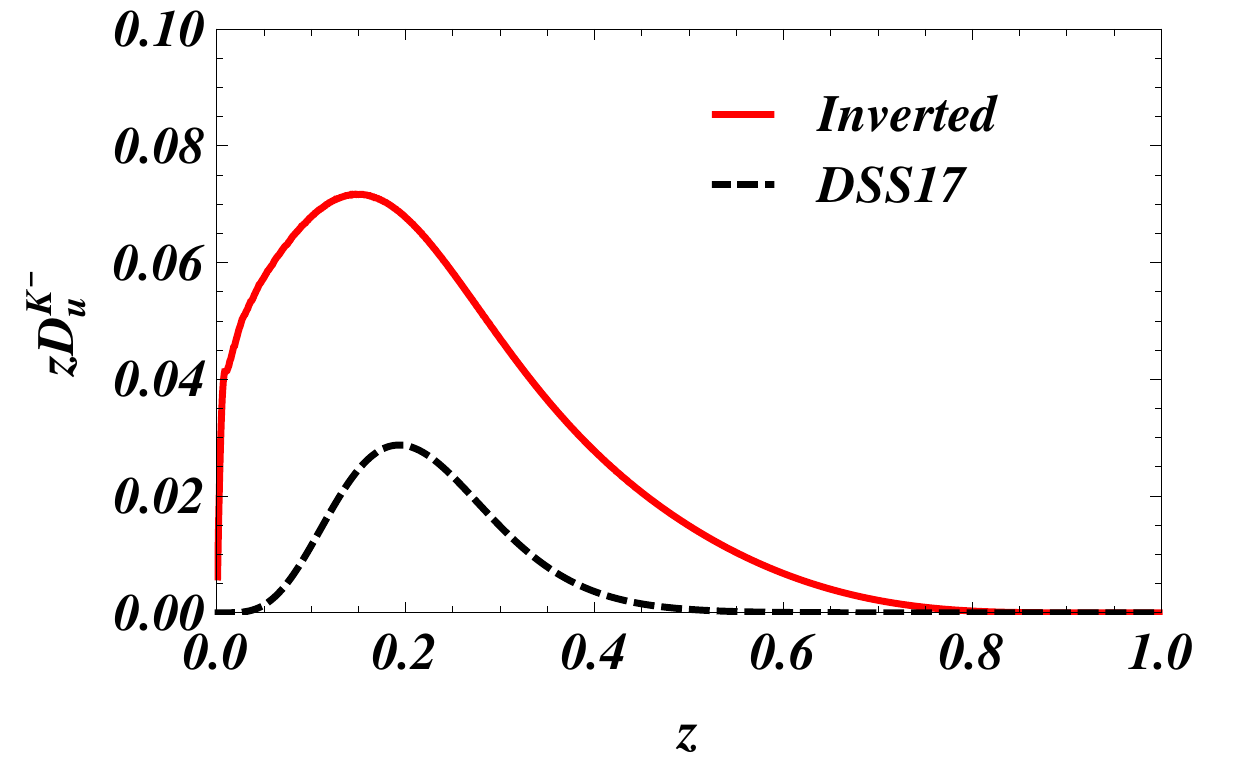}
\includegraphics[width=5.2cm]{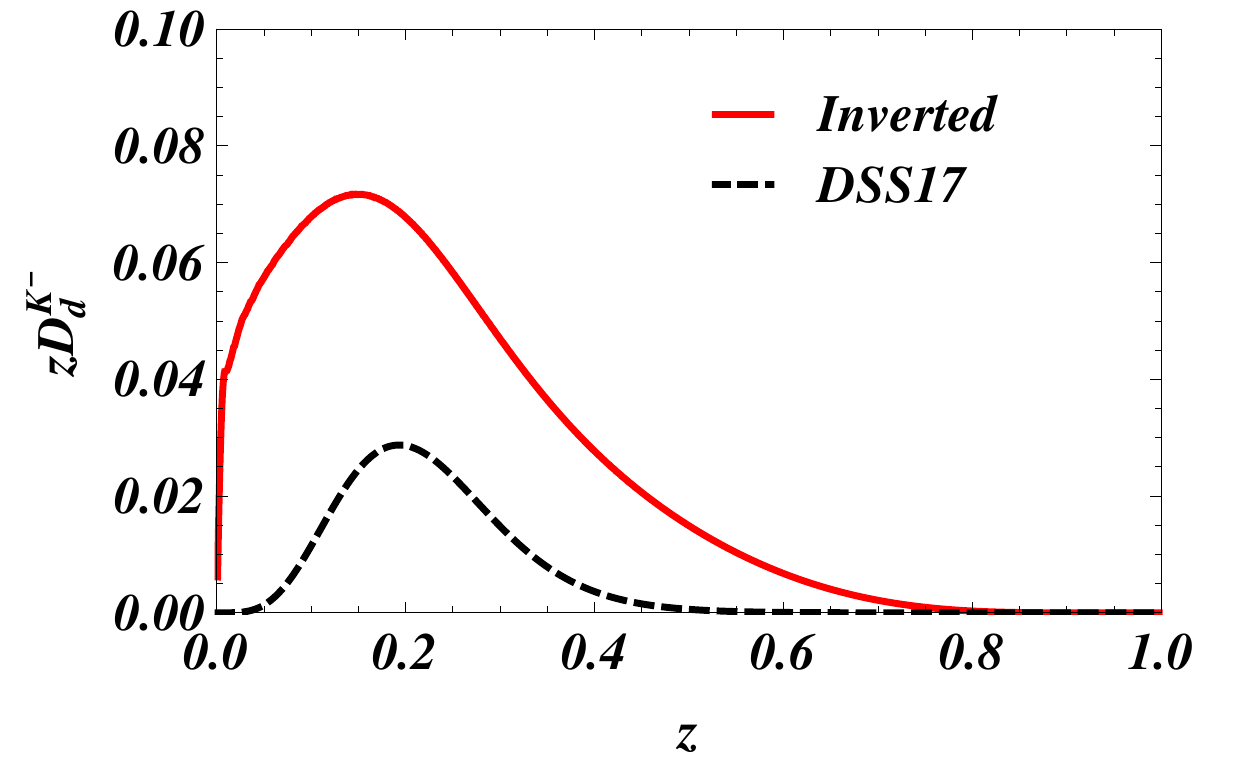}
\end{tabular}
\begin{tabular}{cc}
\includegraphics[width=5.2cm]{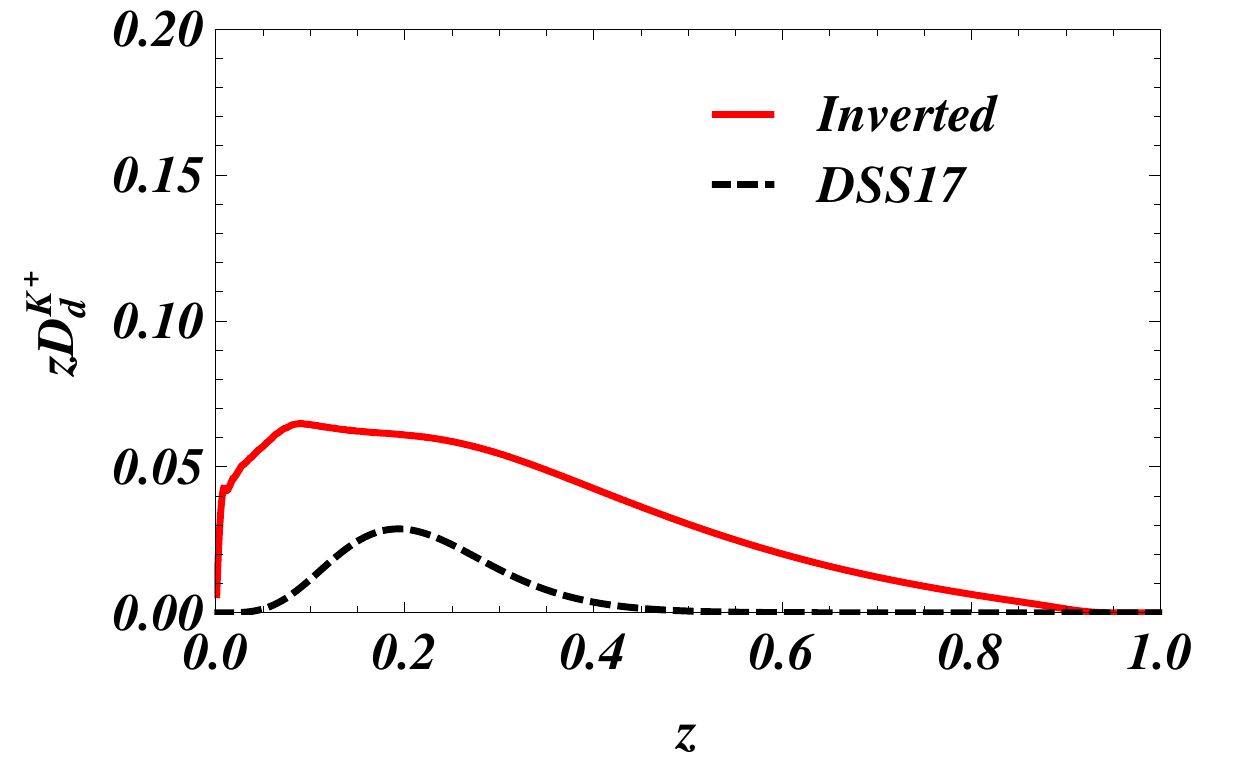}
\includegraphics[width=5.2cm]{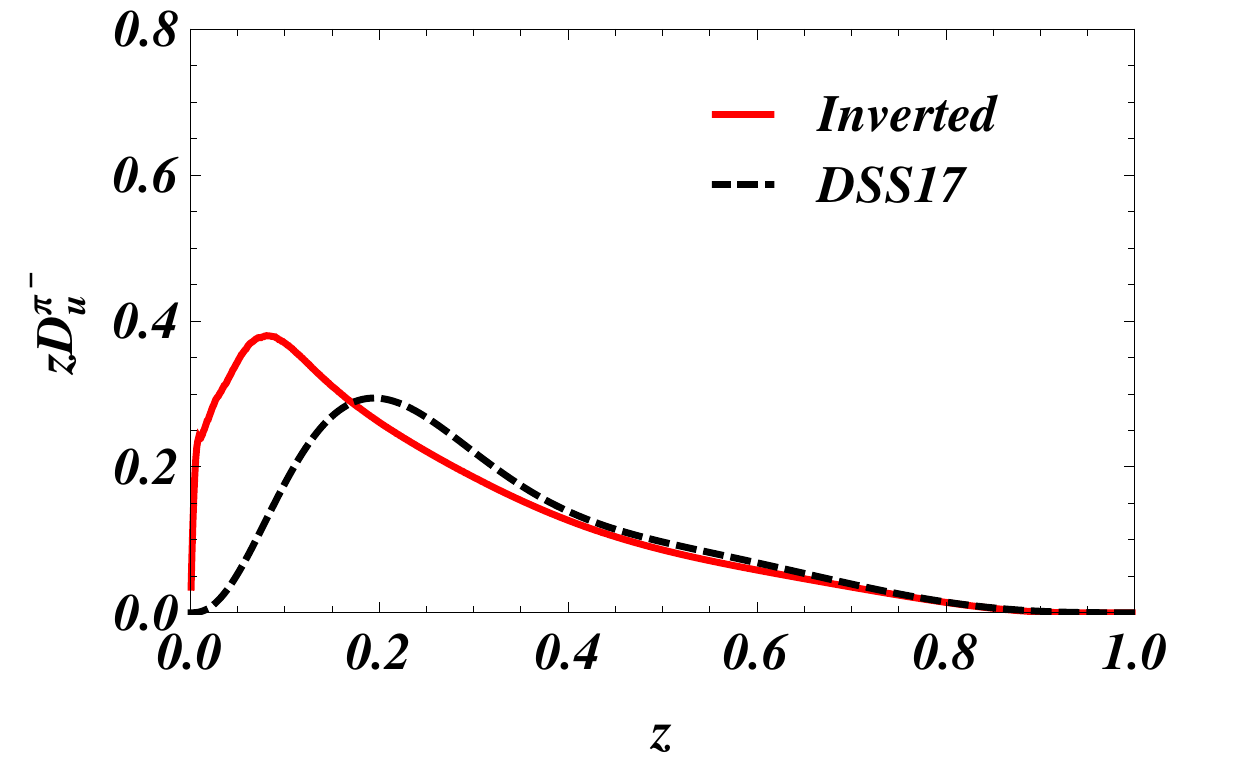}
\end{tabular}
\caption{$zD^{h}_{q}(z)$ at $Q^2=1\,\mathrm{GeV}^2$ for
(1) $(q,h)=(u,K^{-})$=$(d,\bar{K^{0}})$ (left of the top row), (2) $(q,h)=(d,K^{-})$=$(u,\bar{K^{0}})$
(right of the top row), (3) $(q,h)=(d,K^{+})=(u,K^{0})$ (left of the bottom row),
(4)$(q,h)=(u,\pi^{-})$=$(d,\pi^{+})$(right of the bottom row).}
\label{uSFu}
\end{figure}

After this procedure one obtain the results depicted in Fig.~(\ref{curve}). Although this result is much better than the ones in Fig.(~\ref{invert}),
it is still far away from perfect. Therefore we modify this result
by multiplying some arbitrary functions $f_q^m(z)$,
\begin{equation}
d^h_q(z)\equiv d^h_q(z)(iteration)\times f_q^h(z).
\end{equation}
Our choices of $f_q^h$ and the resulting $d^h_q(z)$ are all depicted in Fig~(\ref{zd}).
Using $d_q^h(z)$ presented in Fig~(\ref{zd}) one obtains $D^h_q(z)$ of the favoured channels presented in Fig.~(\ref{FaSiFF}).
We also obtain $D_q^h(z)$ of the unfavoured channels which are demonstrated in Fig.~(\ref{uSFu}).
\\

Our $D^h_q(z)$ for the favoured channels are all very close to the
DSS17 parametrization except
for $D_{u}^{\pi^+}$ in the very low $z$ regime.
As we try to improve our result of $D_{u}^{\pi^+}$, the result of another three
$d_q^h(z)$ become much less satisfactory. Therefore we are satisfied ourselves with the current result.
Our $D^h_q(z)$ for the unfavoured channels of the $u$ and $d$ quarks are presented in Fig.~(\ref{uSFu}).
Our results are in general much larger than the original $D_q^h(z)$ given by the DSS17 parametrization.
In other words, we find that the probabilities of the $u$ and $d$ quarks fragment into the unfavoured kaon
are in general overestimated by the single-cascade jet algorithm.
Fortunately, the magnitude of those SiFFs are actually very small compared with the favoured ones,
therefore this discrepancy is expected to make little effect on the resultant DiFFs.
On the other hand, our result of the fragmentation function of the $u$ quark to $\pi^{-}$
is quite close to the DSS17 parametrization. Its magnitude is much large compared with the kaon ones and it plays more important role
in the coupled-channel calculation of uDiFFs.
\begin{figure}[b]
\begin{tabular}{ccc}
\includegraphics[width=5.2cm]{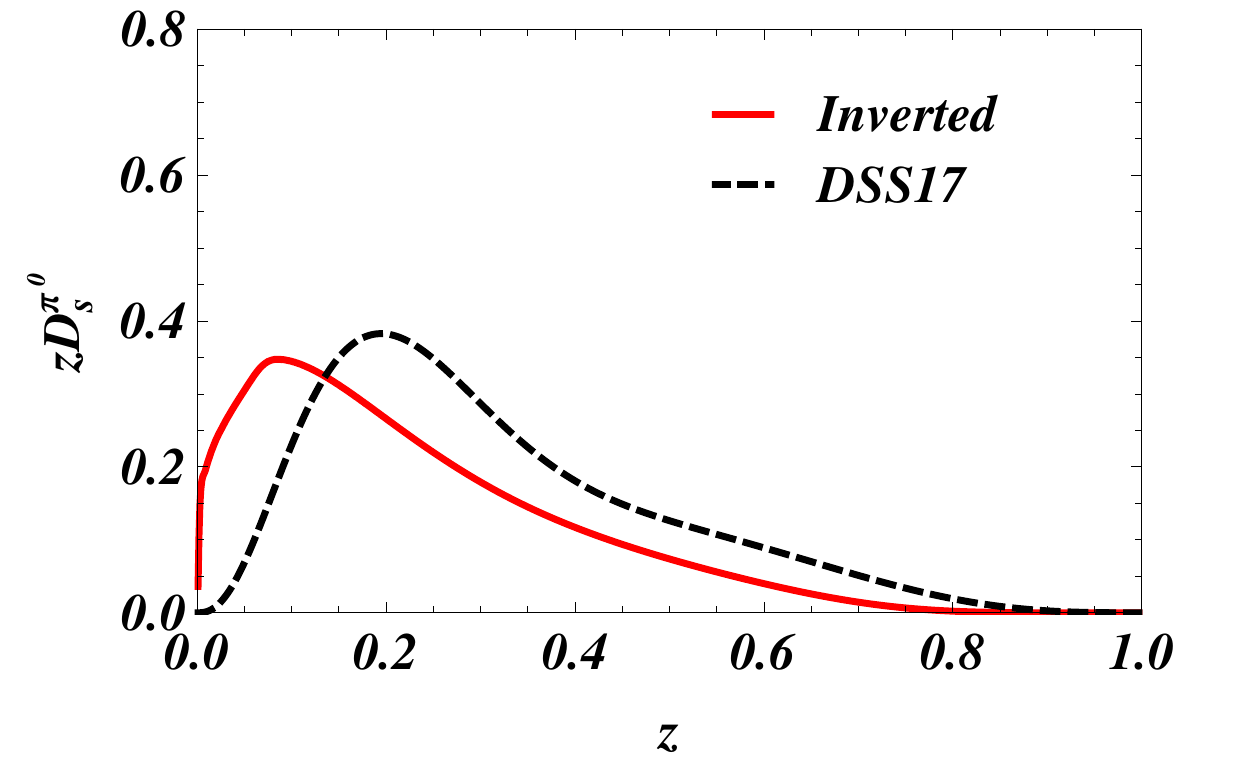}
\includegraphics[width=5.2cm]{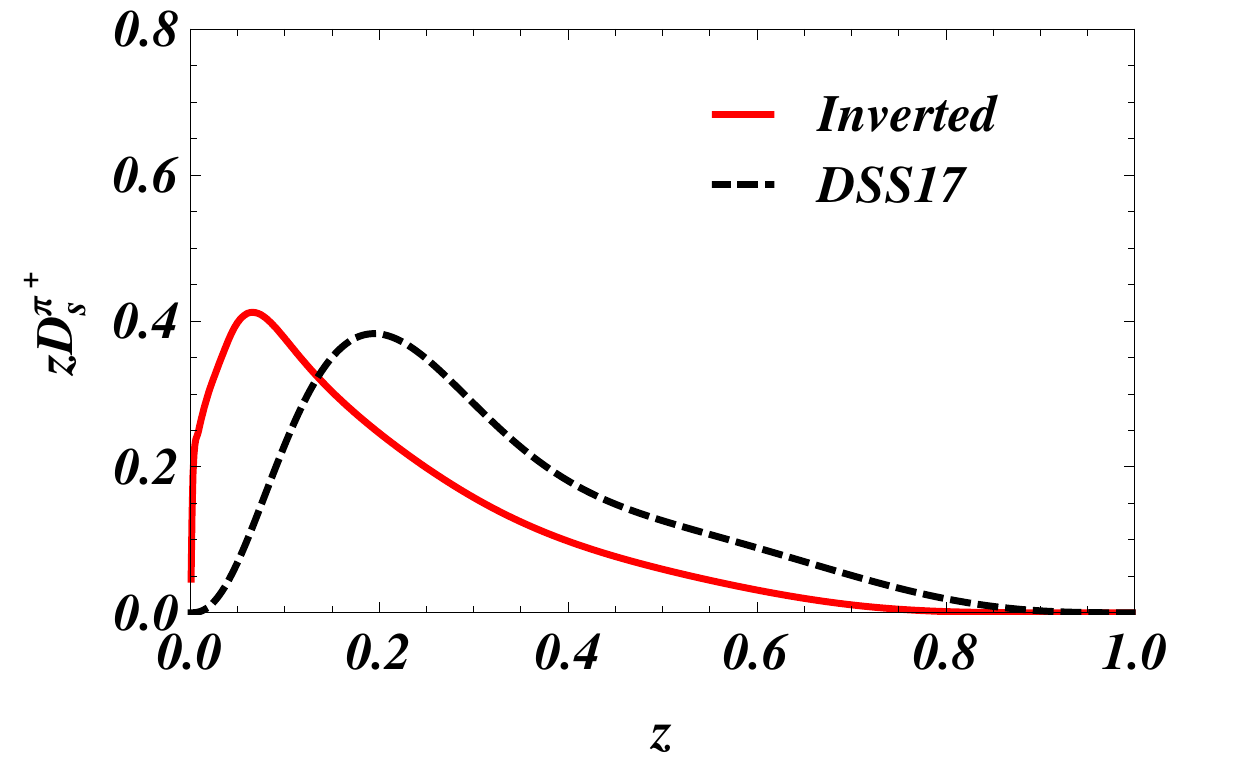}
\includegraphics[width=5.2cm]{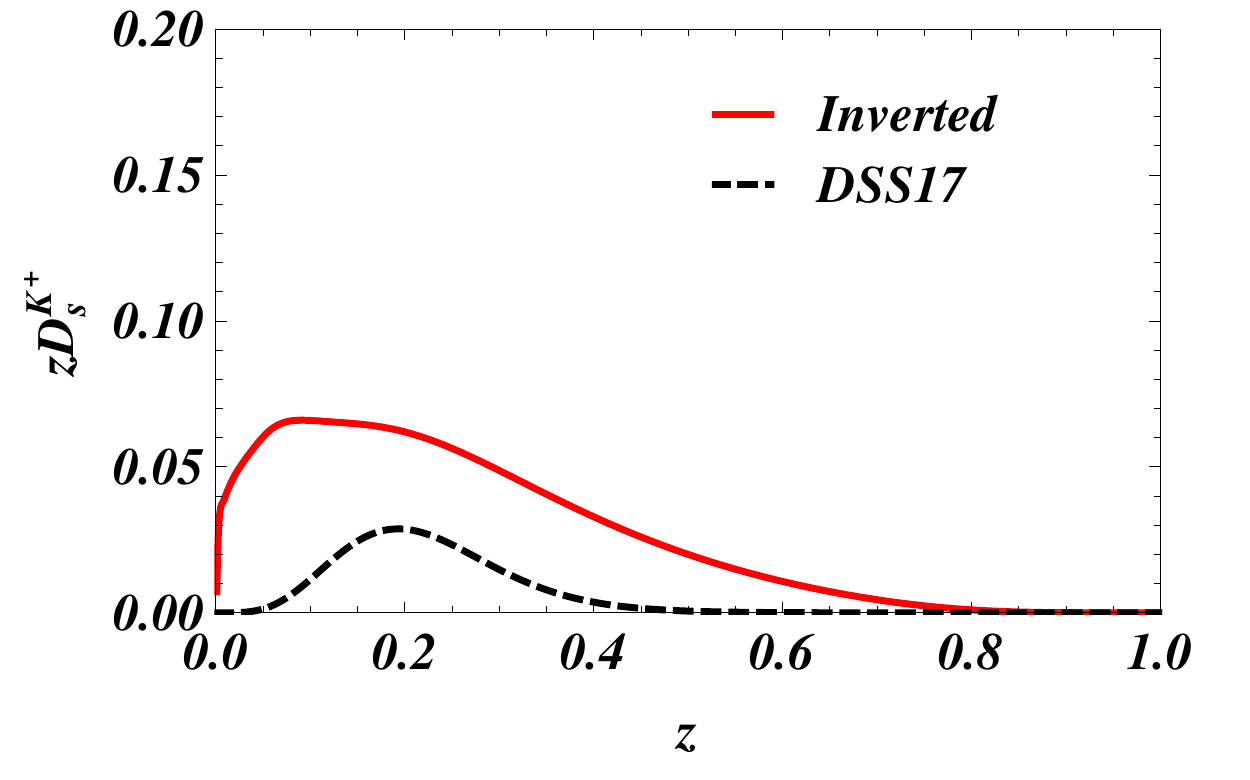}
\end{tabular}
\caption{$zD^{h}_{q}(z)$ at $Q^2=1\,\mathrm{GeV}^2$ for
(1) $(q,h)=(s,\pi^{0})$ (left), (2) $(q,h)=(s,\pi^{+})$=$(s,\pi^{-})$
(middle), (3) $(q,h)=(s,K^{+})$=$(s,K^{0})$ (right).The solid lines are our results generated by our $d^m_q$. The dotted lines are the results
of DSS17 empirical parametrization. }
\label{uSFs}
\end{figure}
Our $D^h_q(z)$ for the unfavoured channels of the strange quark are presented in Fig.~(\ref{uSFs}).
Our result of $D_s^{\pi^{0}}$ and $D_s^{\pi^{+}}$ are quite close to the
results of the DSS17 parametrizations with the $z$ value shifted to the left about $0.1$.
The magnitude of these two SiFFs are about the same with the other favoured SiFFs.
Our result of $D_s^{K^{+}}(z)$ is two times larger than the DSS17 ones, however this channel is just $10\%$ of the another ones.
Another interesting observation is that all results of the unfavoured channels of the kaon in our approach are larger than the DSS17 results about two to three
times. Fortunately their magnitude is smaller than the favoured ones substantially so they are not expected to generate large deviations.
Nevertheless this trend itself shows that the approach beyond the single-cascade jet algorithm will be required for the further study of the fragmentation functions.
\section{Result and discussion}

With the new elementary fragmentation functions $d^h_q(z)$ presented in Fig.~(\ref{zd}) and the resultant single-hadron fragmentation functions $D^h_q(z)$ presented in
Fig.~(\ref{zd},~\ref{uSFu},~\ref{uSFs}), it is straightforward to obtain the dihadron fragmentation functions by the following equations,
\begin{equation}
D^{h_{1},h_{2}}_{q}(z_{1},z_{2})=\delta_{aq}\hat{d}^{h_{1}}_{q}(z_{1})\frac{D^{h_{2}}_{q_{1}}\left(\frac{z_{2}}{1-z_{1}}\right)}{1-z_{1}}+
\delta_{bq}\hat{d}^{h_{2}}_{q}(z_{2})\frac{D^{h_{1}}_{q_{2}}\left(\frac{z_{1}}{1-z_{2}}\right)}{1-z_{2}}
+\sum_{Q}\int^{1}_{z_{1}+z_{2}}\frac{d\eta}{\eta^2}\hat{d}^{Q}_{q}(\eta)D^{h_{1},h_{2}}_{Q}\left(\frac{z_{1}}{\eta},\frac{z_2}{\eta}\right).
\label{Dihadron1}
\end{equation}
Here the flavour component of the emitted hadrons $h_1$ and $h_2$ are
$h_{1}=(a\bar{q}_{1})$ and $h_{2}=(b\bar{q}_{2})$, respectively.
When $q$ is neither $a$ nor $b$
then $D^{h_1,h_2}_{q}(z_1,z_2)$ is called the disfavored uDiFF. Otherwise it is called the favored uDiFF.
The first term stands for the situation that $h_1$ is the first emitted hadron in the decay cascade of the quark $q$.
Similarly the second term denotes the situation that $h_2$ is the first emitted hadron.
The last term represents the situation that the first emitted hadron is neither $h_{1}$ nor $h_{2}$.
To simplify the equation we choose new valuables
as $\xi_{1}=z_{1}/\eta$ and $\xi_{2}=z_2/\eta$:
\begin{eqnarray}
D^{h_{1},h_{2}}_{q}(z_{1},z_{2})&=&\delta_{qa}\hat{d}^{h_{1}}_{q}(z_{1})\frac{D^{h_{2}}_{q_{1}}\left(\frac{z_{2}}{1-z_{1}}\right)}{1-z_{1}}+
\delta_{qb}\hat{d}^{h_{2}}_{q}(z_{2})\frac{D^{h_{1}}_{q_{2}}\left(\frac{z_{1}}{1-z_{2}}\right)}{1-z_{2}} \nonumber \\
&+&\sum_{Q}\int^{\frac{z_1}{z_{1}+z_{2}}}_{z_{1}}d\xi_{1}\int^{\frac{z_1}{z_1+z_2}}_{z_2}d\xi_{2}\delta(z_2\xi_{1}-z_{1}\xi_{2})
\hat{d}^{Q}_{q}(z_{1}/\xi_{1})D^{h_{1},h_{2}}_{Q}(\xi_{1},\xi_{2}).
\label{Dihadron2}
\end{eqnarray}

\begin{figure}[t]
\begin{tabular}{ccc}
\includegraphics[width=5.2cm]{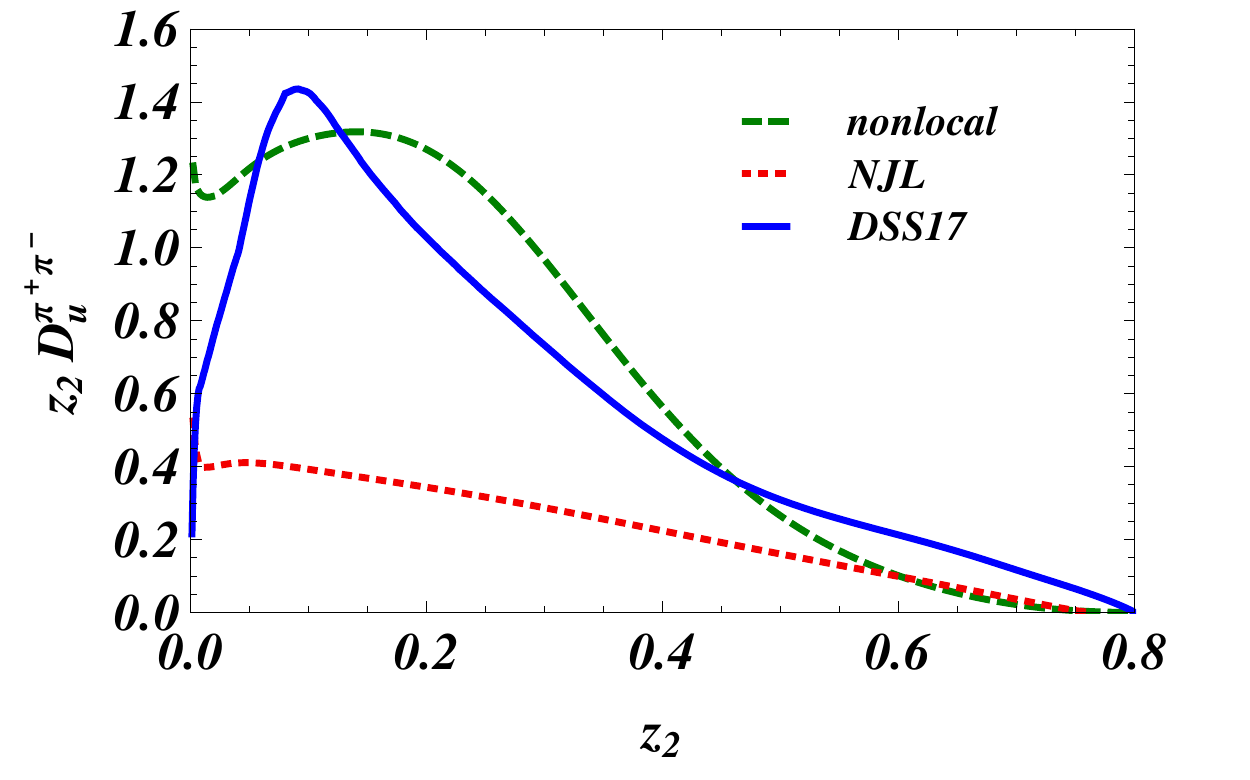}
\includegraphics[width=5.2cm]{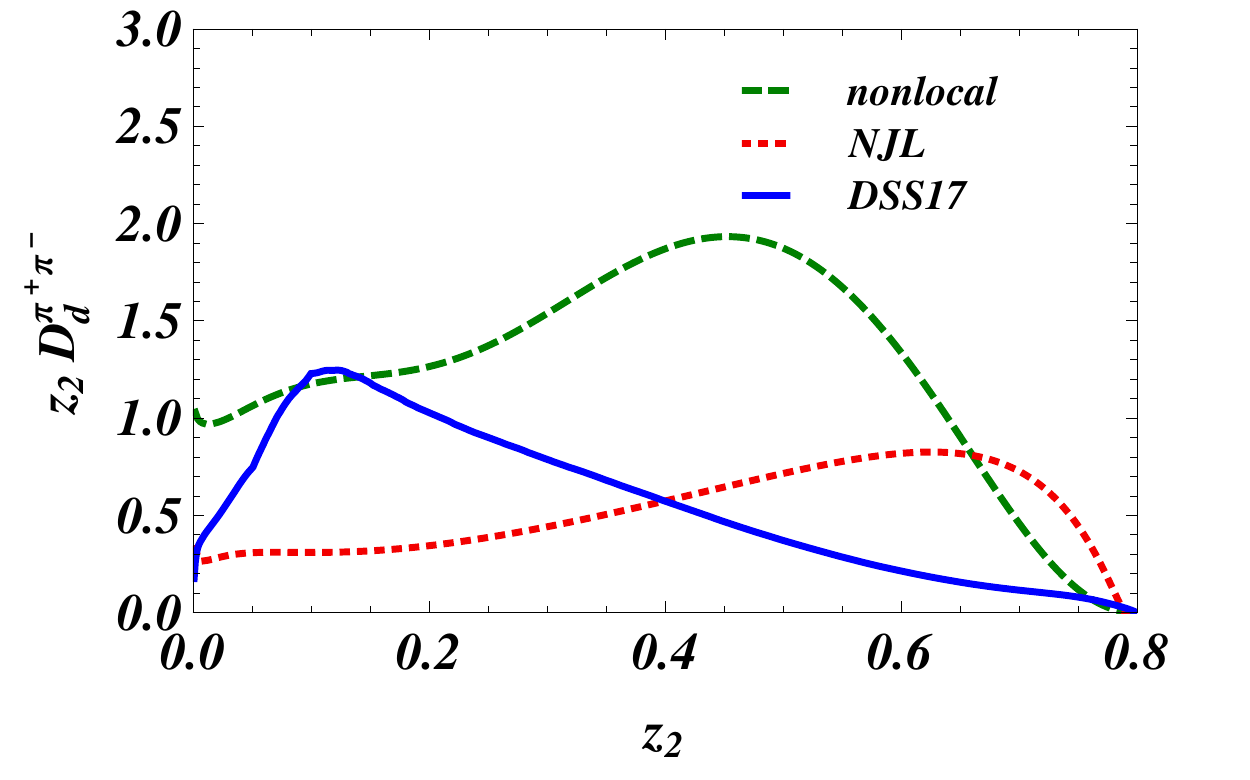}
\includegraphics[width=5.2cm]{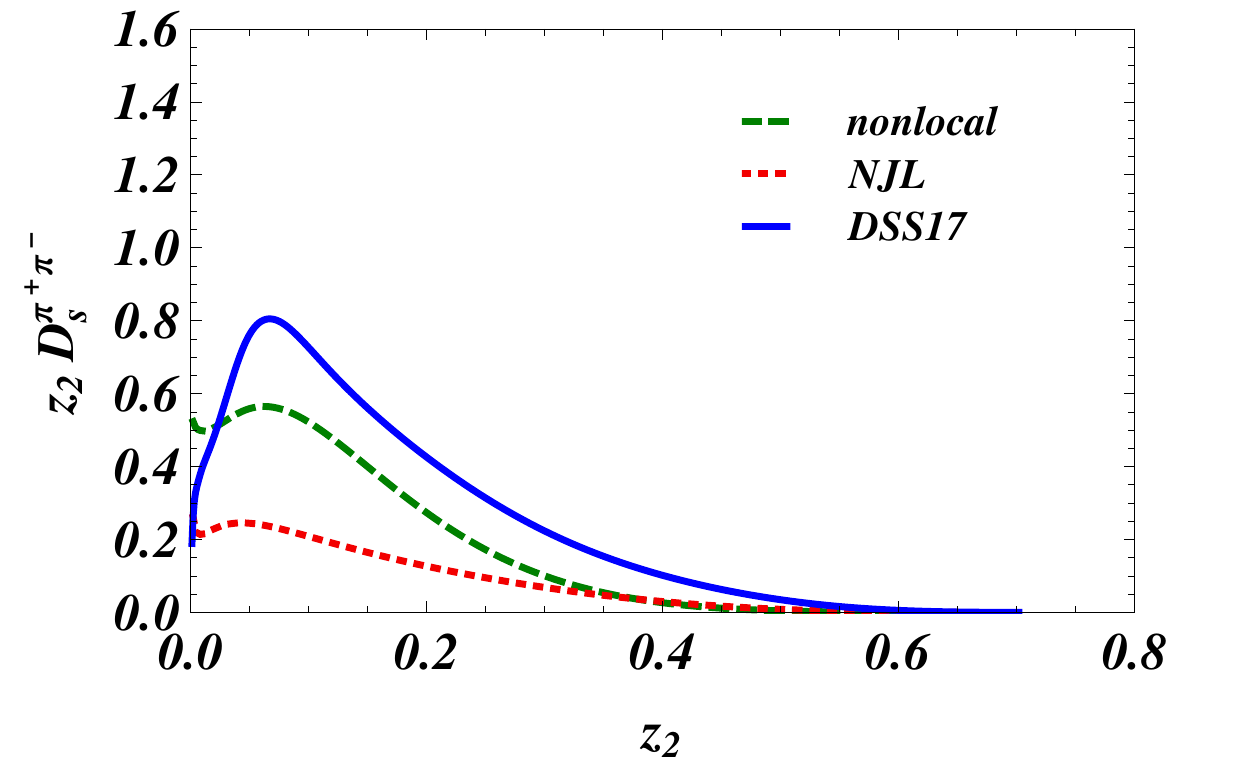}
\end{tabular}
\begin{tabular}{ccc}
\includegraphics[width=5.2cm]{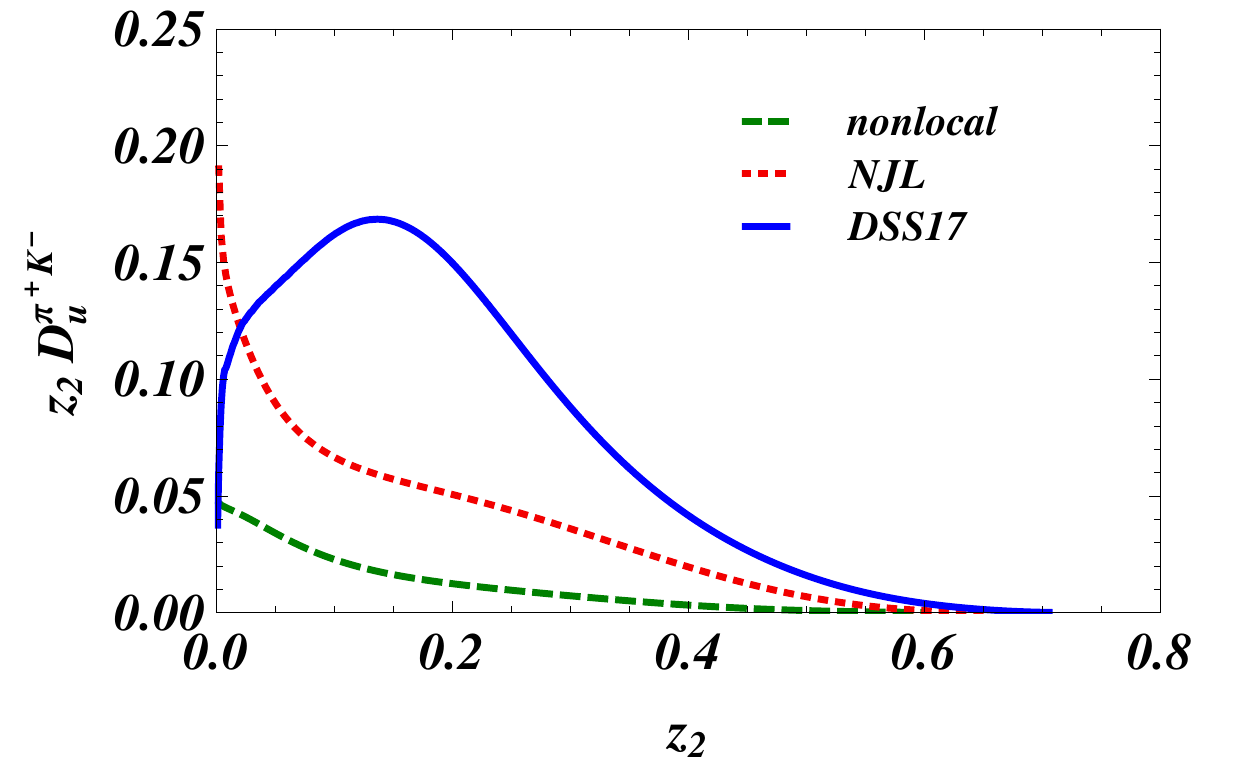}
\includegraphics[width=5.2cm]{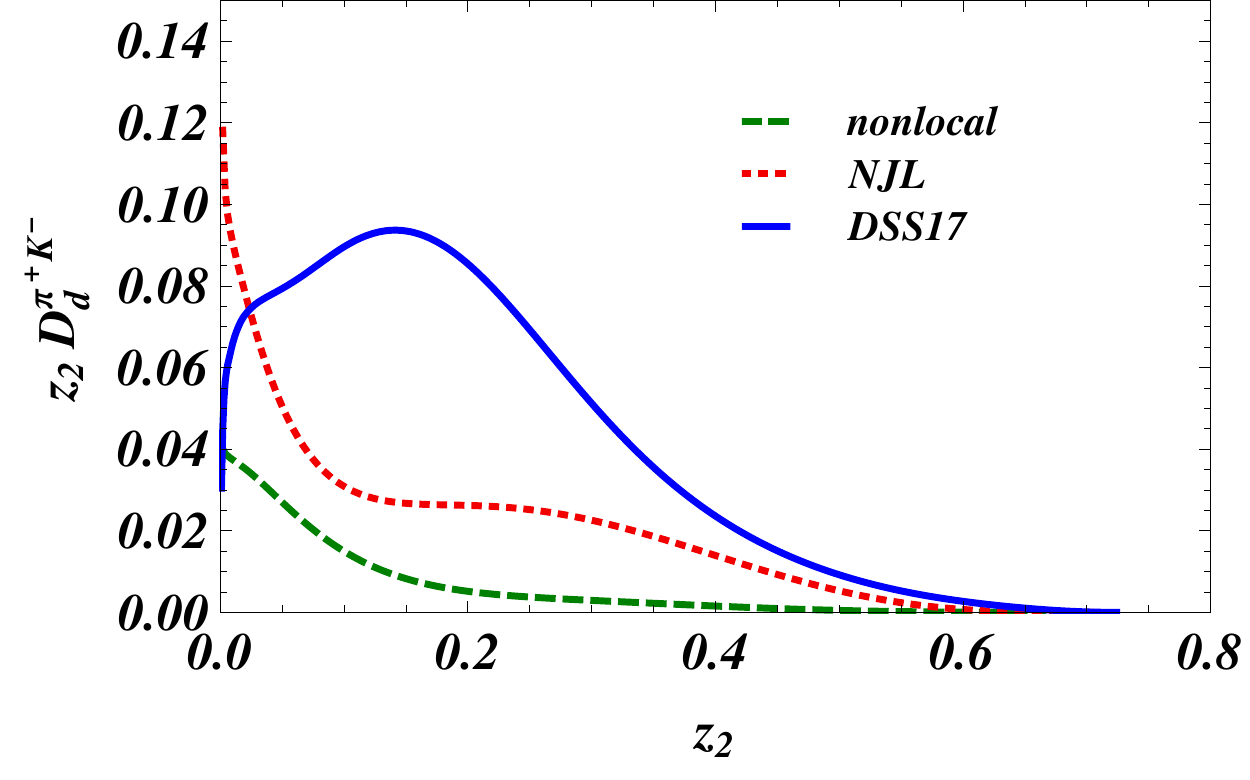}
\includegraphics[width=5.2cm]{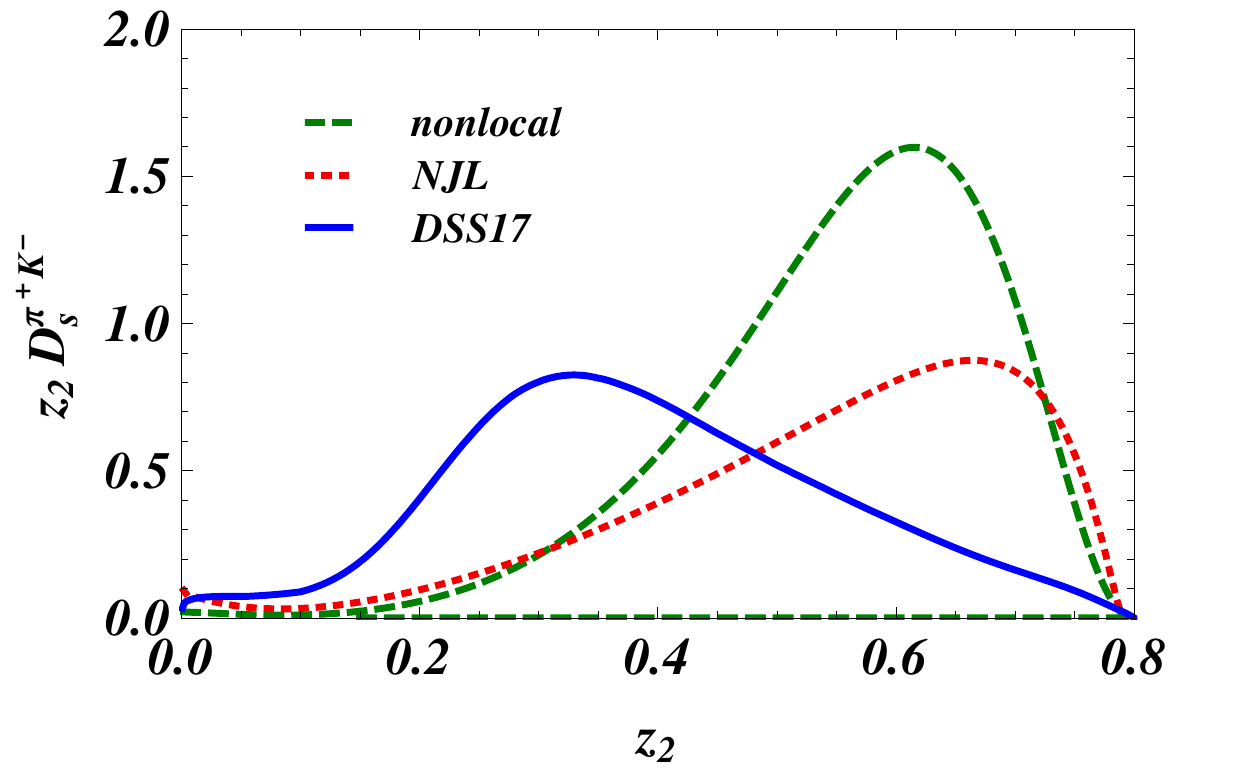}
\end{tabular}
\begin{tabular}{ccc}
\includegraphics[width=5.2cm]{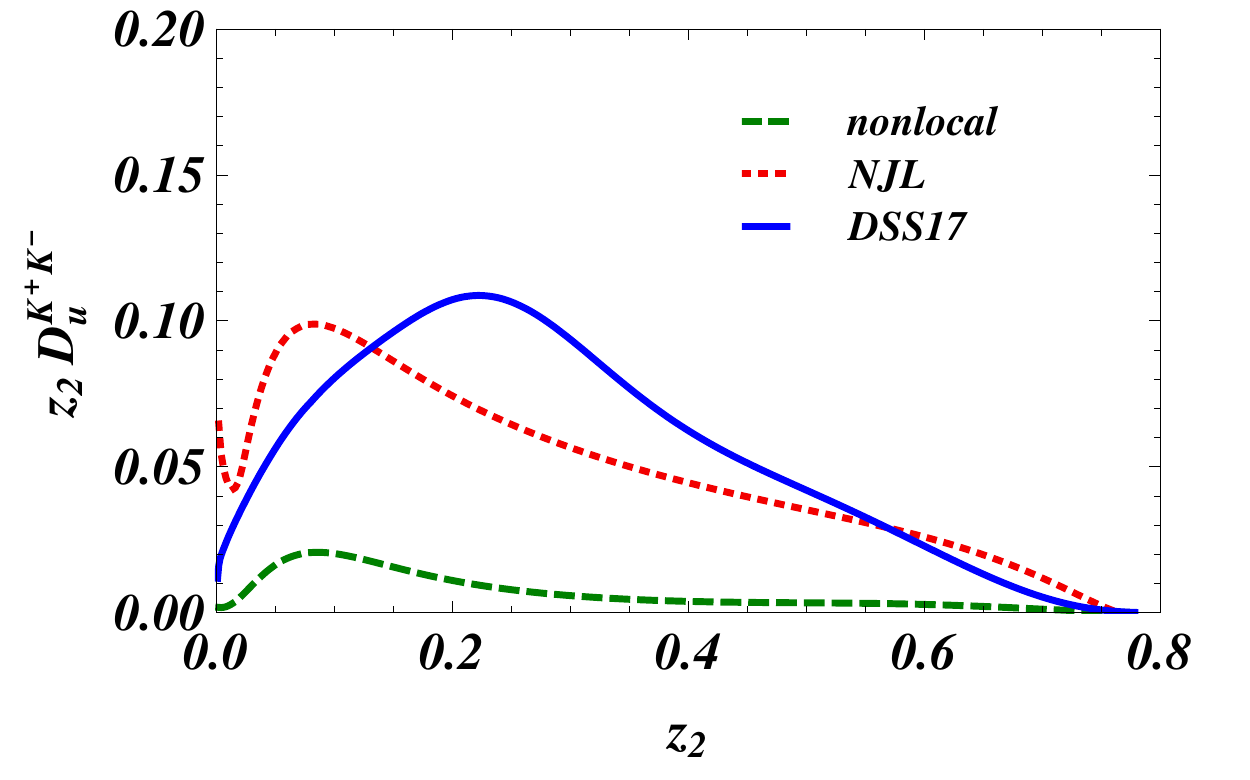}
\includegraphics[width=5.2cm]{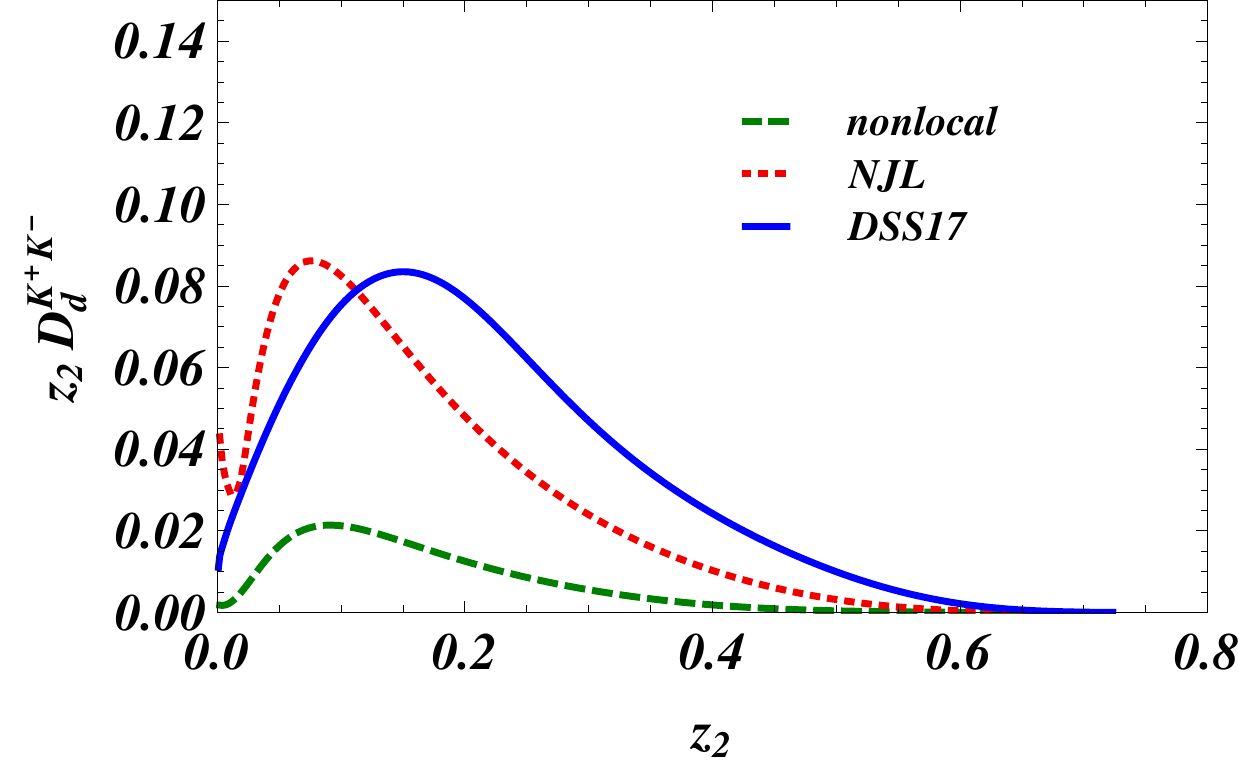}
\includegraphics[width=5.2cm]{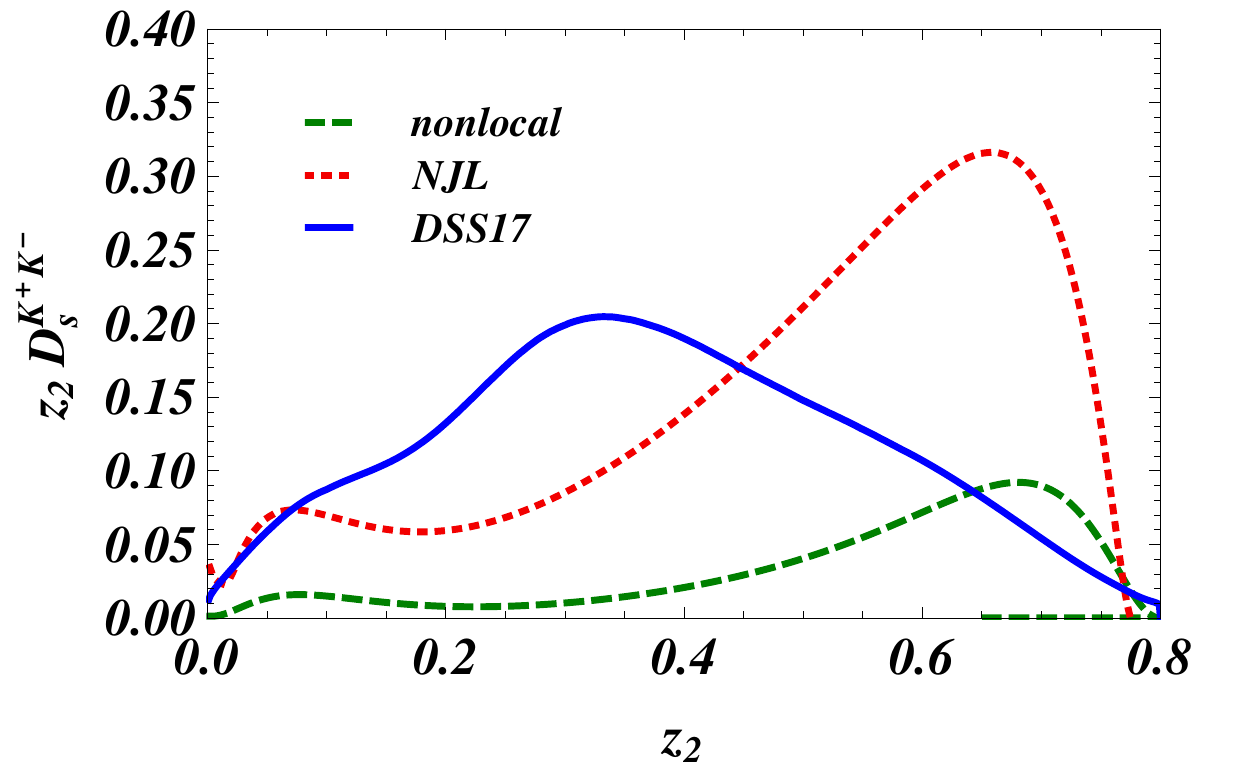}
\end{tabular}
\caption{$z_2 D^{h_1,h_2}_{q}(z_1,z_2)$ with $z_{1}=0.2$ and $Q^2=4\,\mathrm{GeV}^2$ for
(1) $(q,h_1,h_2)=(u,\pi^{+},\pi^{-})$ (left of the top row), (2) $(d,\pi^{+},\pi^{-})$
(middle of the top row), (3) $(s,\pi^{+},\pi^{-})$ (right of the top row),
(4) $(u,\pi^{+},K^{-})$ (left of the middle row), (5) $(d,\pi^{+},K^{-})$
(middle of the middle row), (6) $(s,\pi^{+},K^{-})$ (right of the middle row)
(7) $(u,K^{+},K^{-})$ (left of the bottom row), (8) $(d,K^{+},K^{-})$
(middle of the bottom row), (9) $(s,K^{+},K^{-})$ (right of the bottom row).
The dashed and solid lines denote the results of the NJL-jet model and the nonlocal chiral quark model respectively.
The range of $z_2$ is from zero to 0.8.}
\label{uDi2}
\end{figure}

\begin{figure}
\begin{tabular}{ccc}
\includegraphics[width=5.2cm]{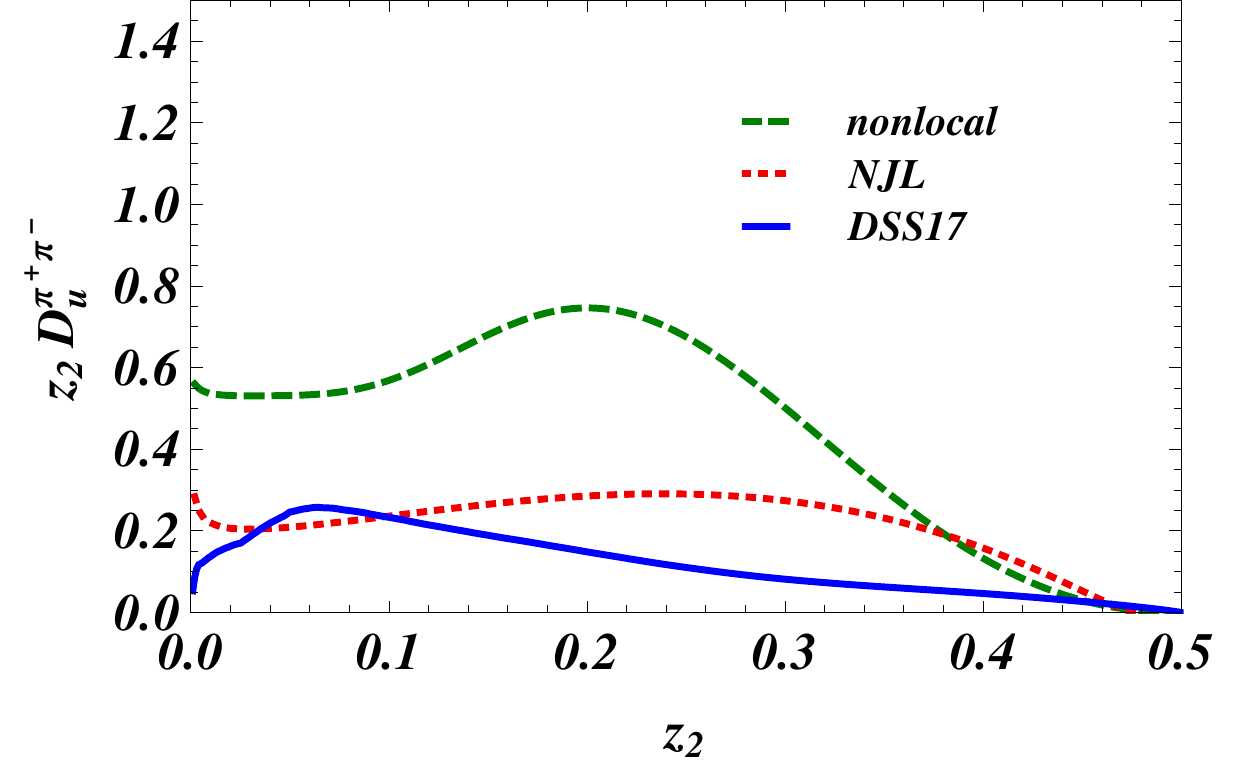}
\includegraphics[width=5.2cm]{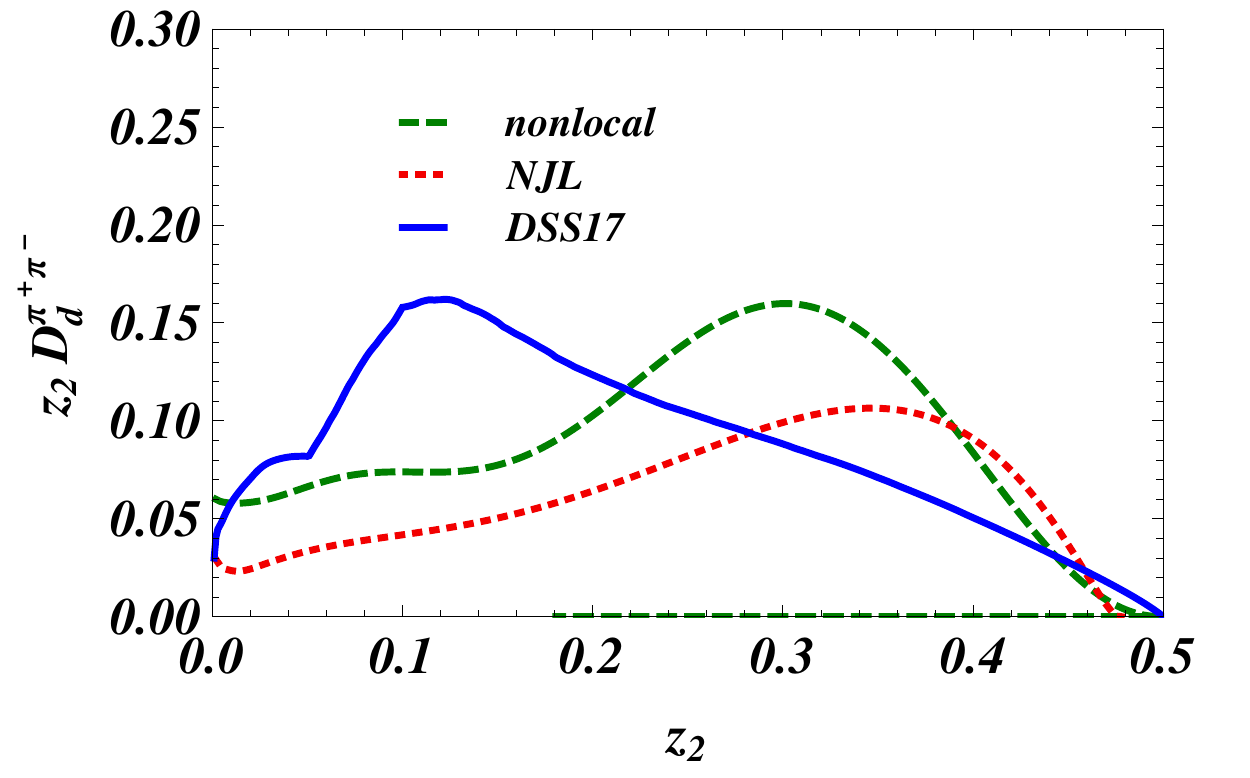}
\includegraphics[width=5.2cm]{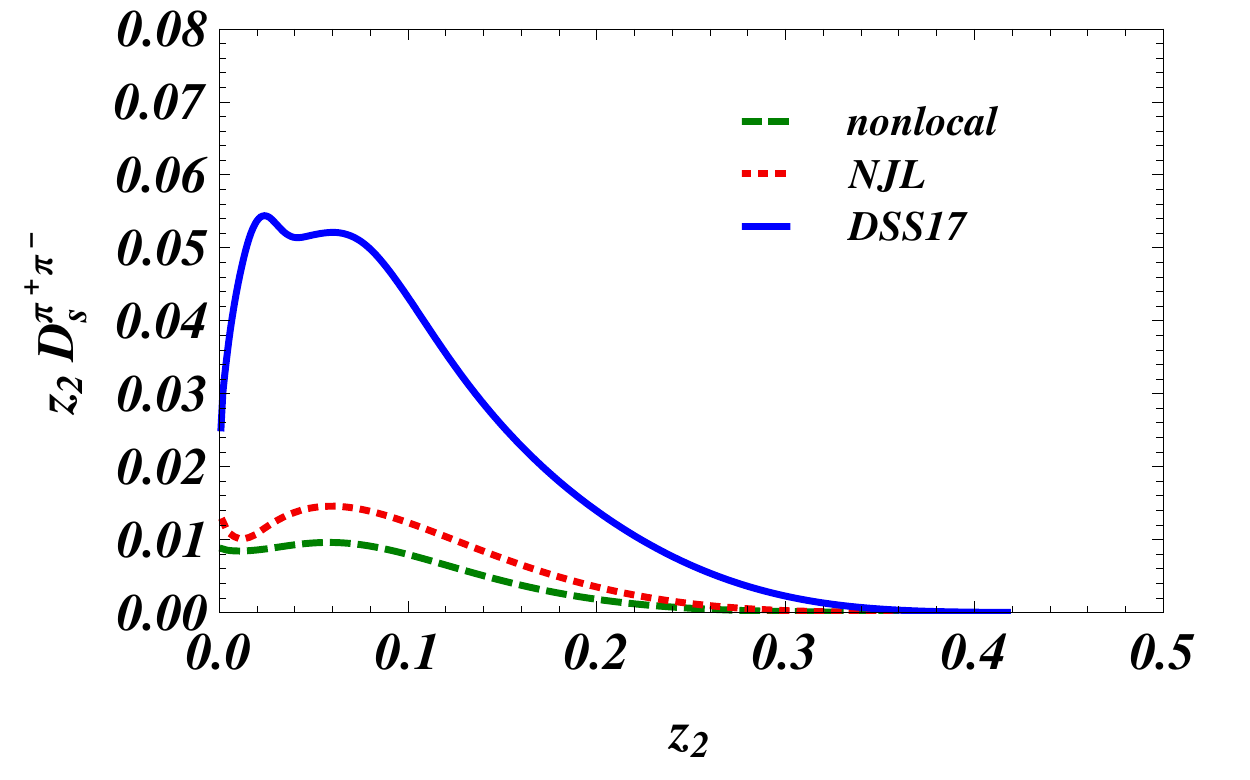}
\end{tabular}
\begin{tabular}{ccc}
\includegraphics[width=5.2cm]{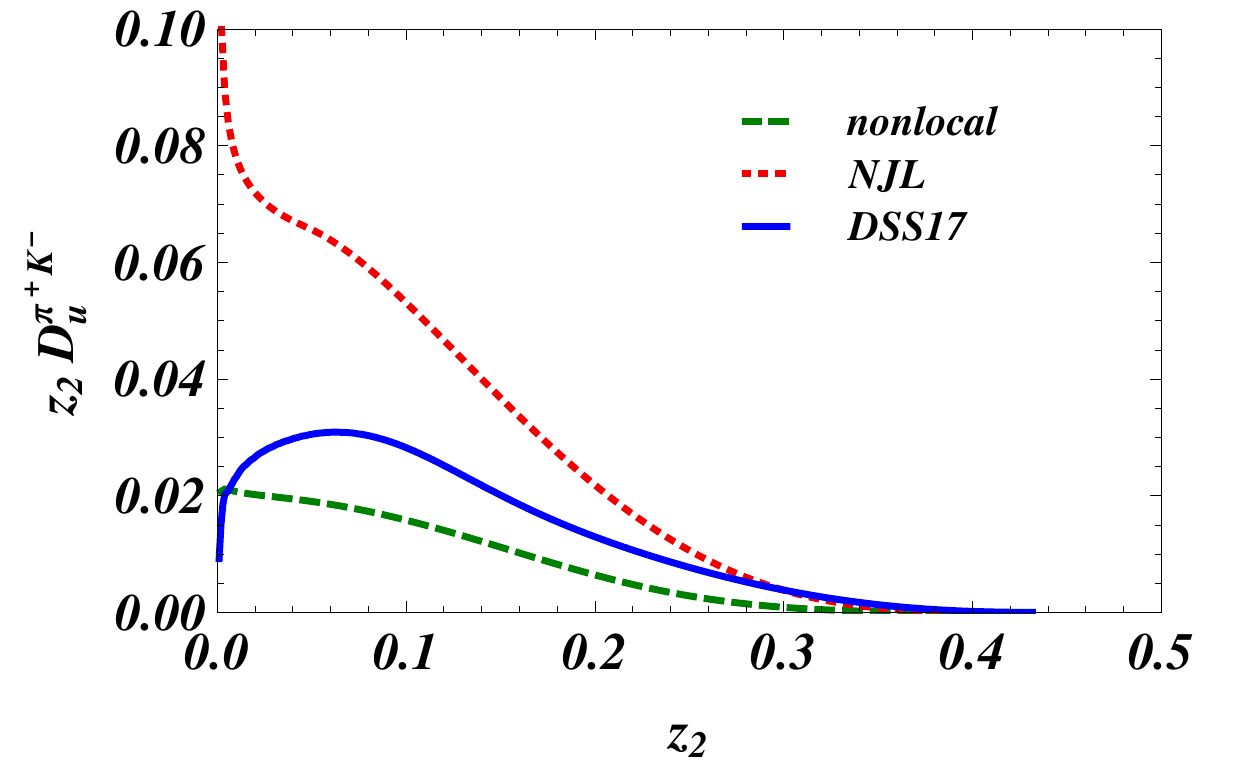}
\includegraphics[width=5.2cm]{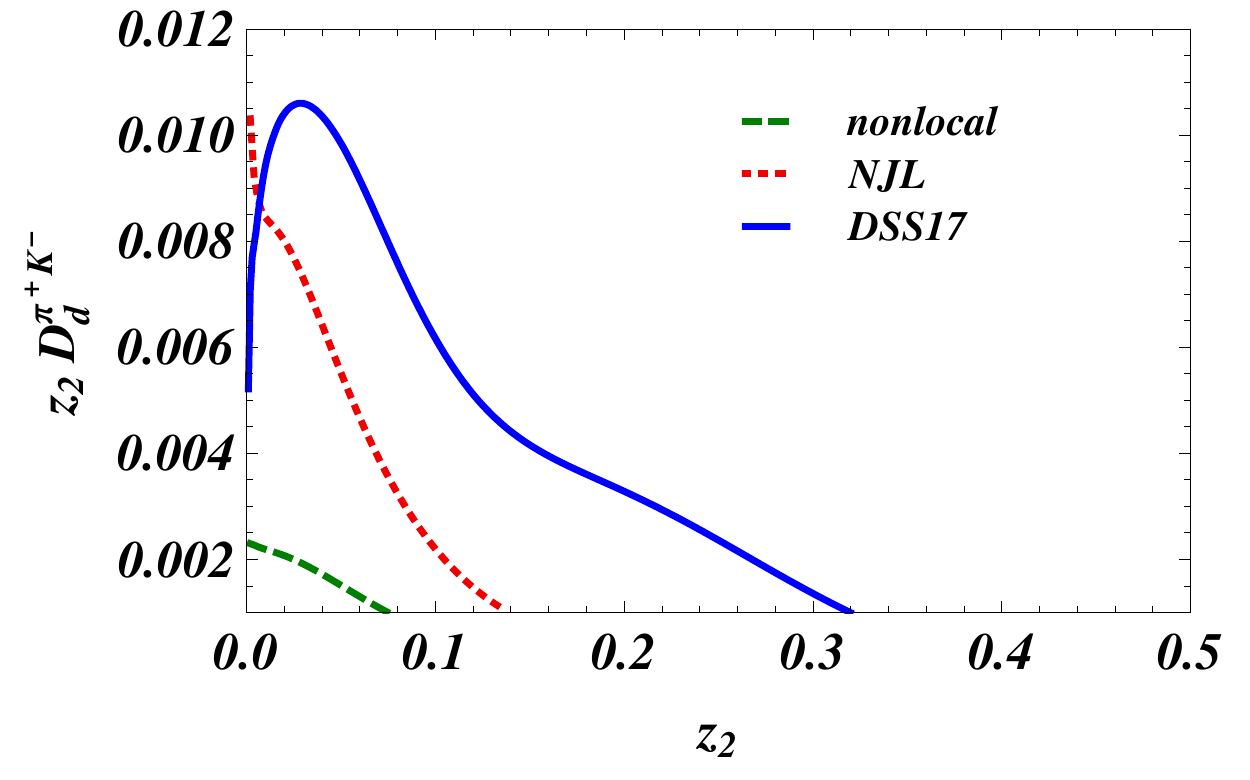}
\includegraphics[width=5.2cm]{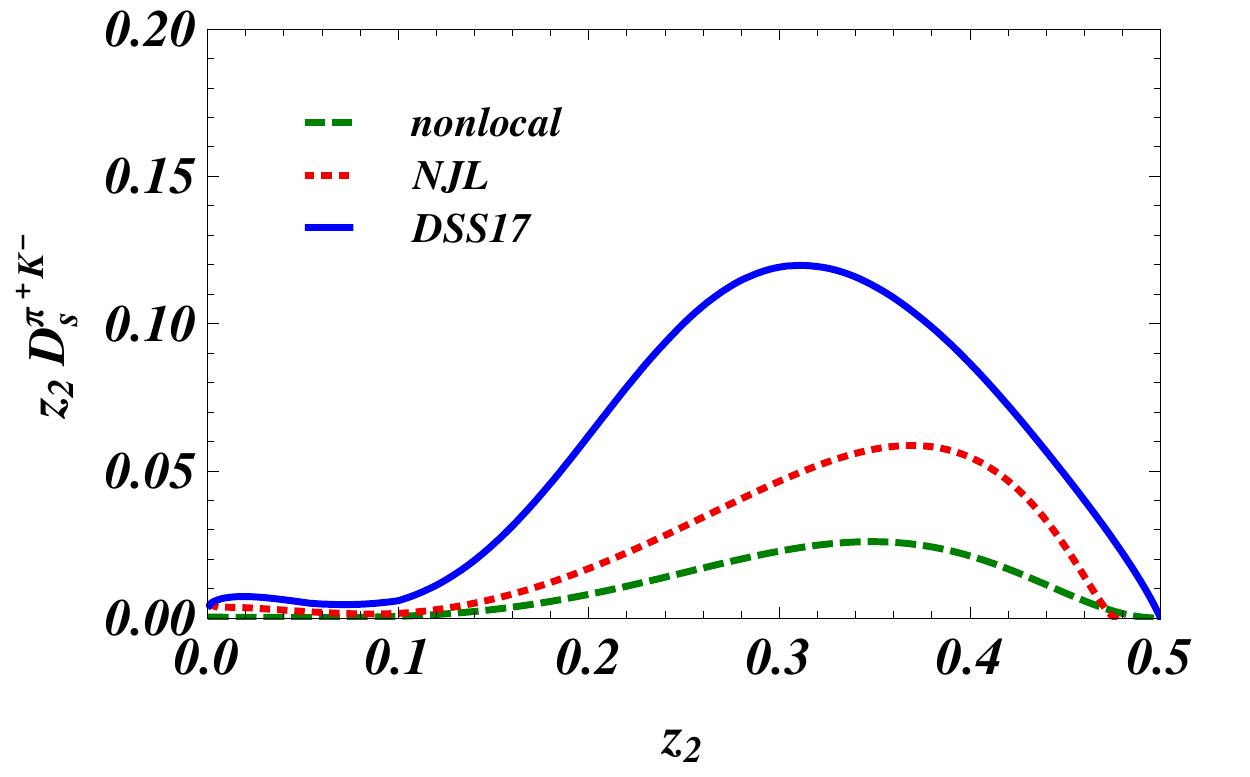}
\end{tabular}
\begin{tabular}{ccc}
\includegraphics[width=5.2cm]{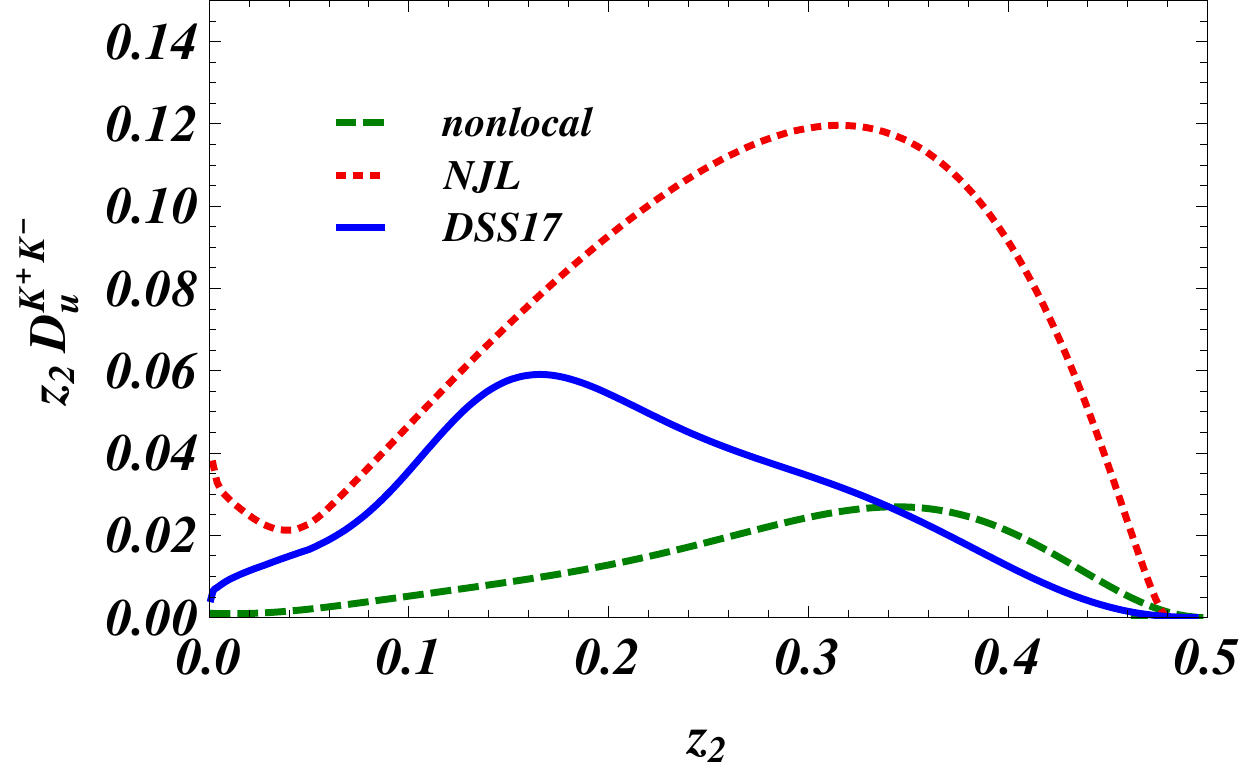}
\includegraphics[width=5.2cm]{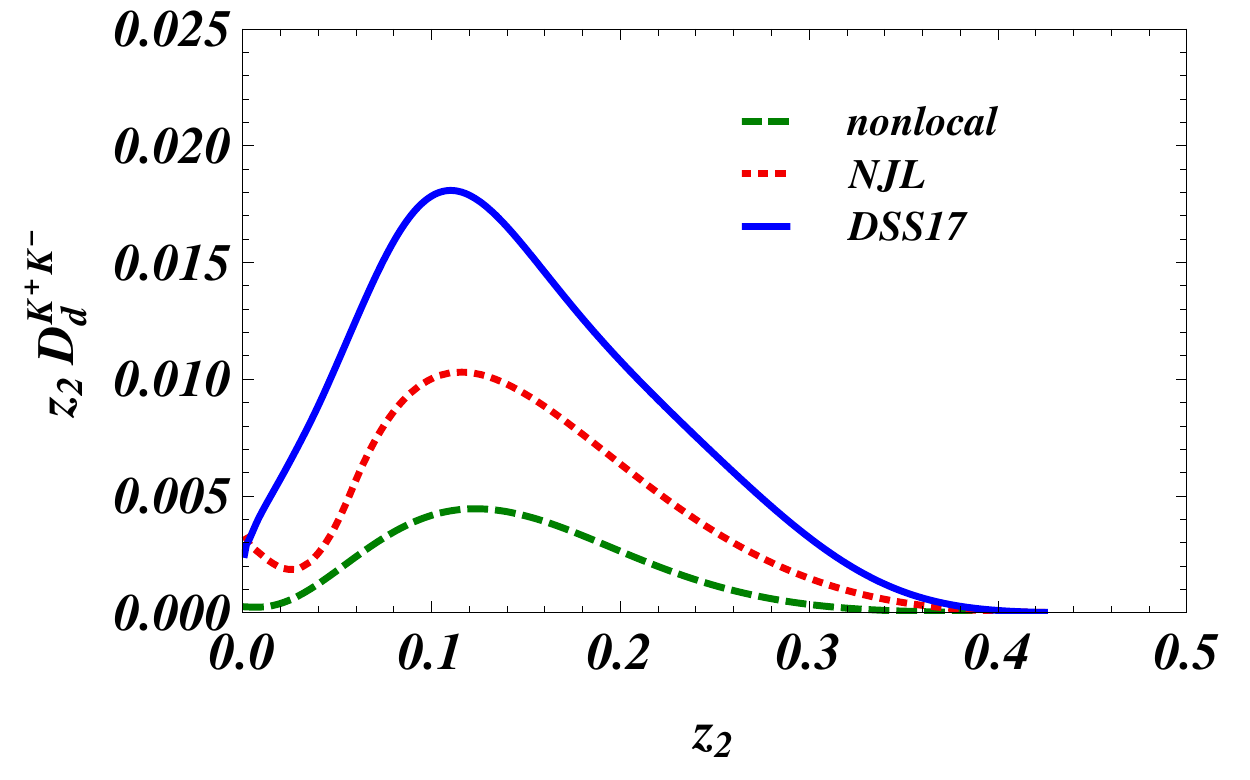}
\includegraphics[width=5.2cm]{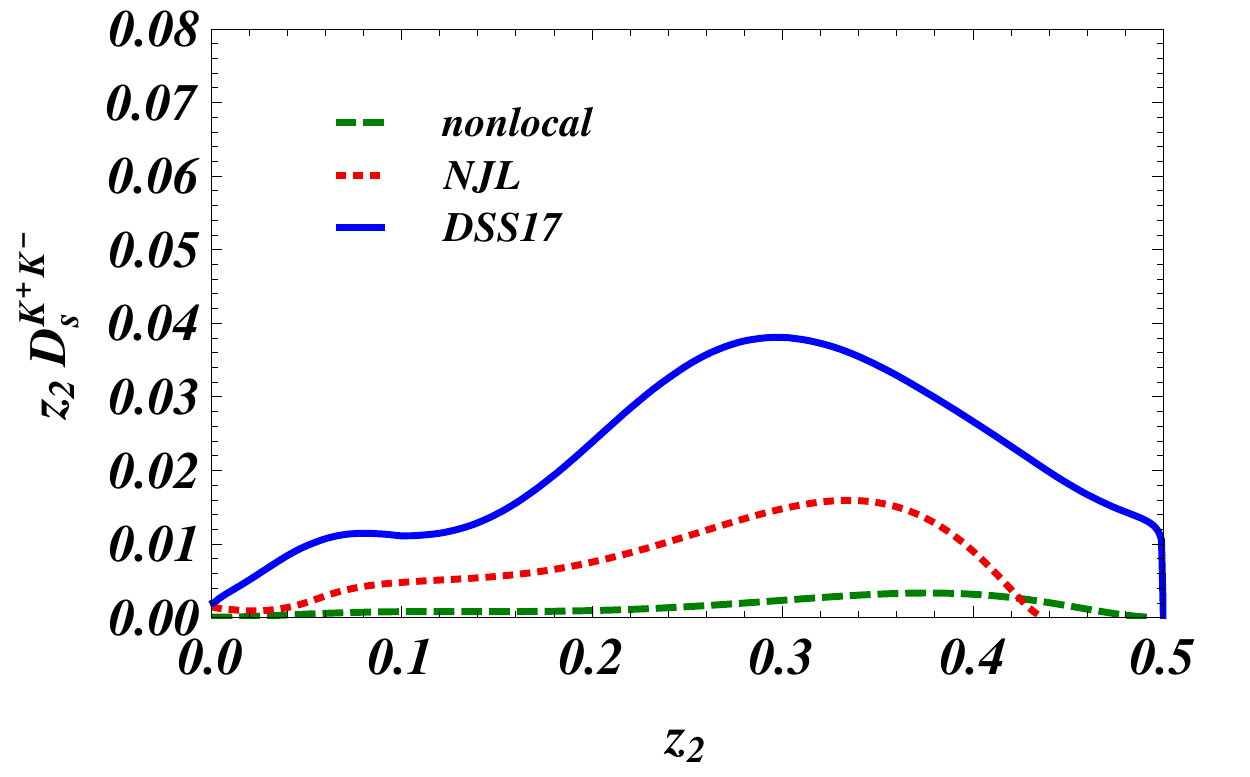}
\end{tabular}
\caption{$z_2 D^{h_1,h_2}_{q}(z_1,z_2)$ with $z_{1}=0.5$ and $Q^2=4\,\mathrm{GeV}^2$ for
(1) $(q,h_1,h_2)=(u,\pi^{+},\pi^{-})$ (left of the top row), (2) $(d,\pi^{+},\pi^{-})$
(middle of the top row), (3) $(s,\pi^{+},\pi^{-})$ (right of the top row),
(4) $(u,\pi^{+},K^{-})$ (left of the middle row), (5) $(d,\pi^{+},K^{-})$
(middle of the middle row), (6) $(s,\pi^{+},K^{-})$ (right of the middle row)
(7) $(u,K^{+},K^{-})$ (left of the bottom row), (8) $(d,K^{+},K^{-})$
(middle of the bottom row), (9) $(s,K^{+},K^{-})$ (right of the bottom row).
The dashed and solid lines denote the results of the NJL-jet model and the nonlocal chiral quark model respectively. The range of $z_2$ is from zero to 0.5.
}
\label{uDi5}
\end{figure}

\begin{figure}
\begin{tabular}{ccc}
\includegraphics[width=5.2cm]{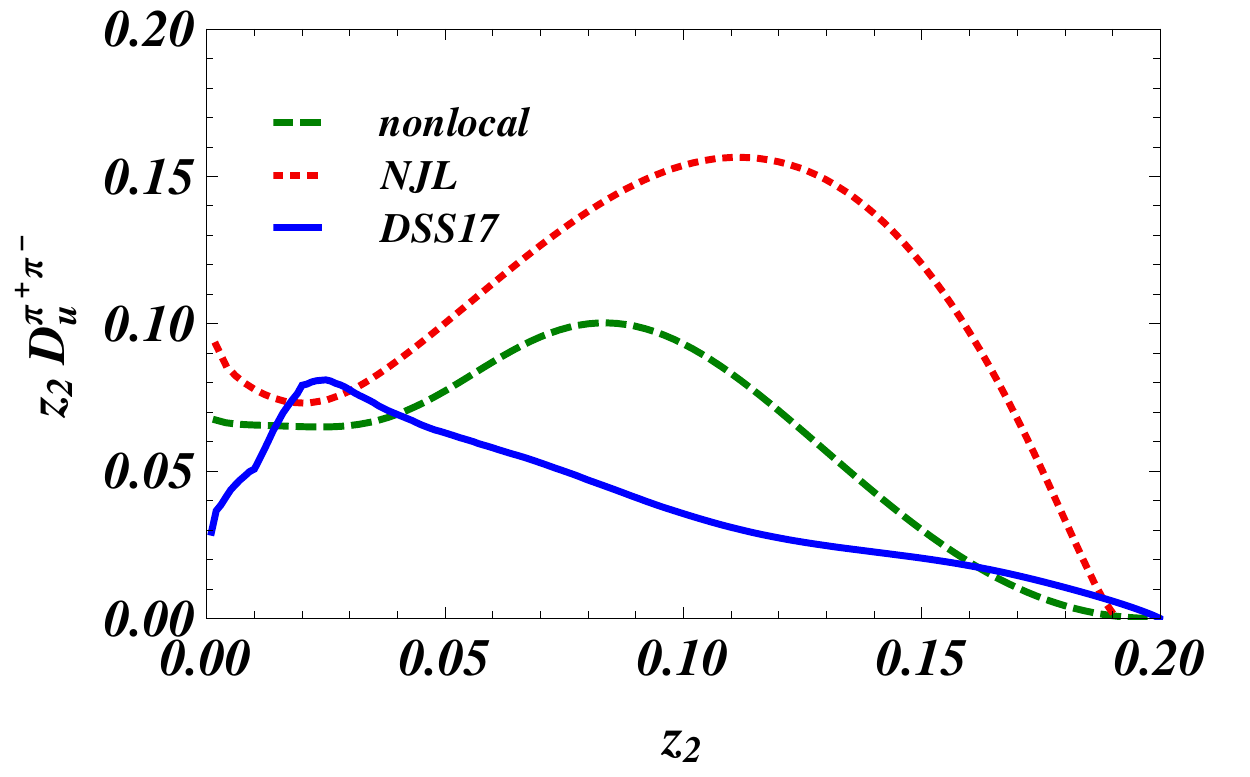}
\includegraphics[width=5.2cm]{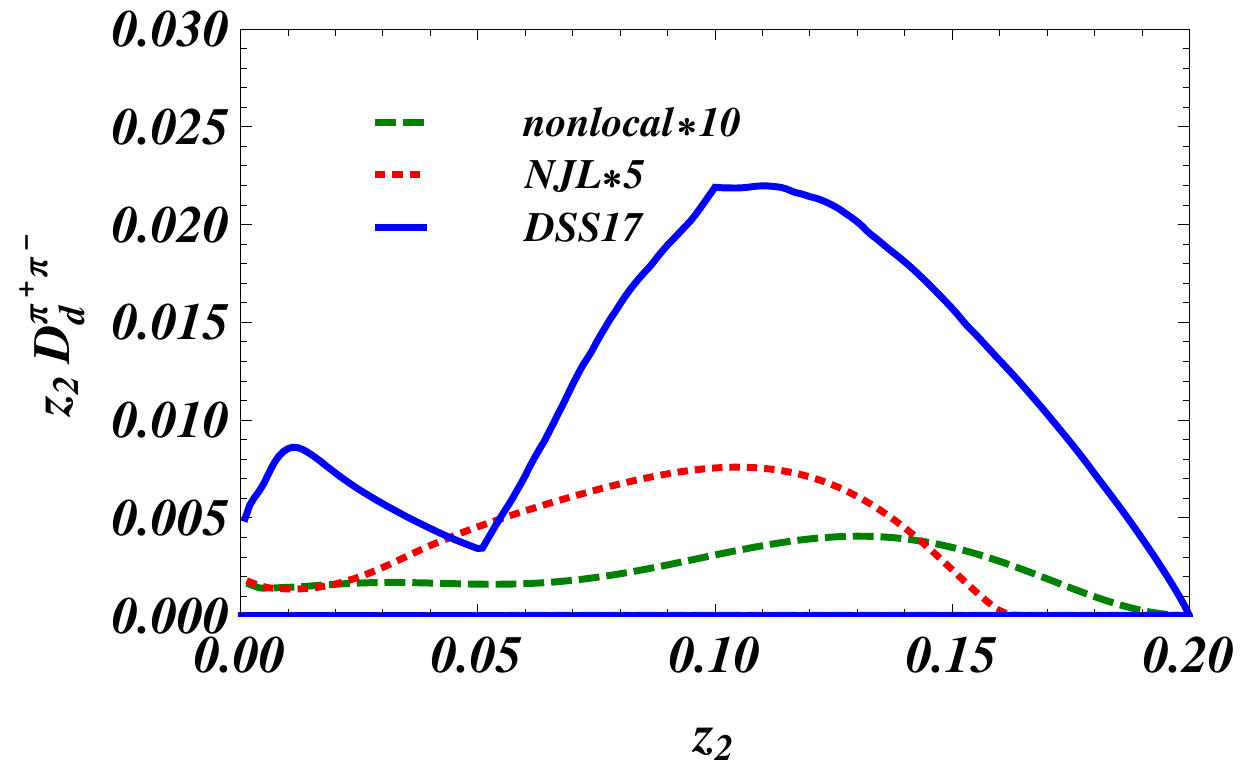}
\includegraphics[width=5.2cm]{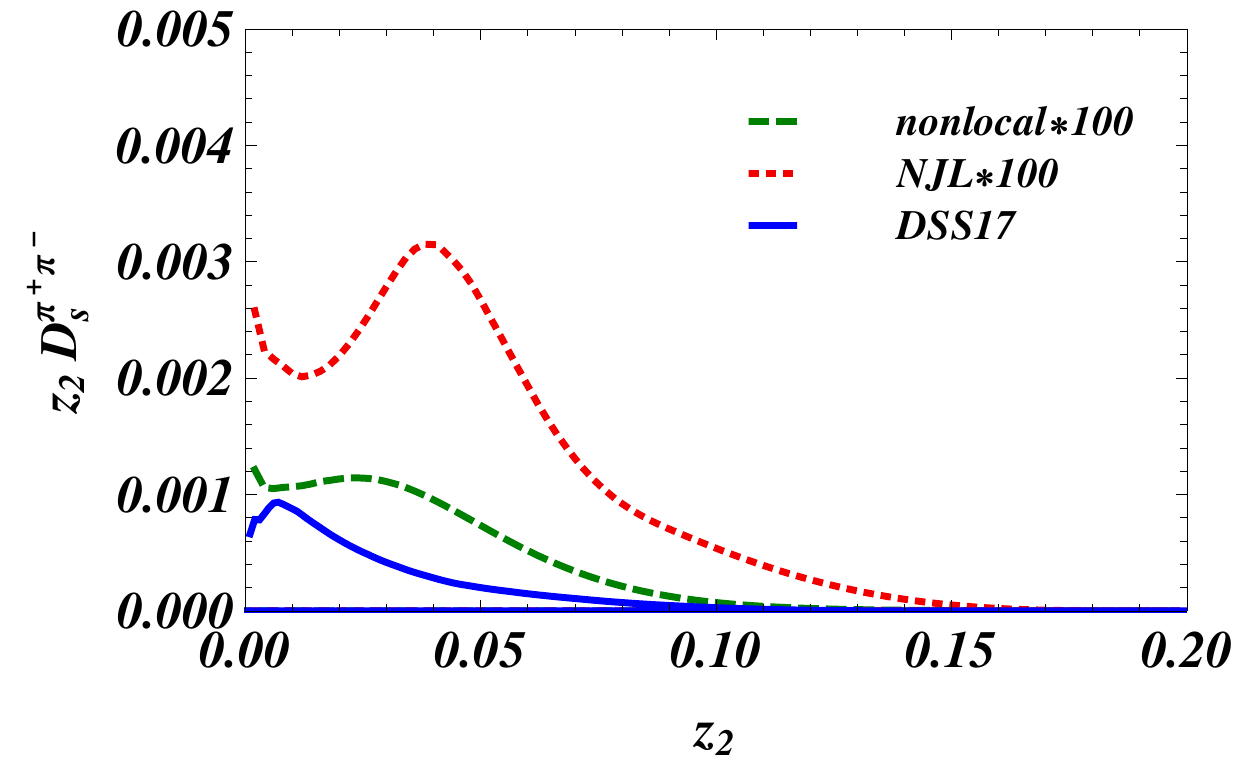}
\end{tabular}
\begin{tabular}{ccc}
\includegraphics[width=5.2cm]{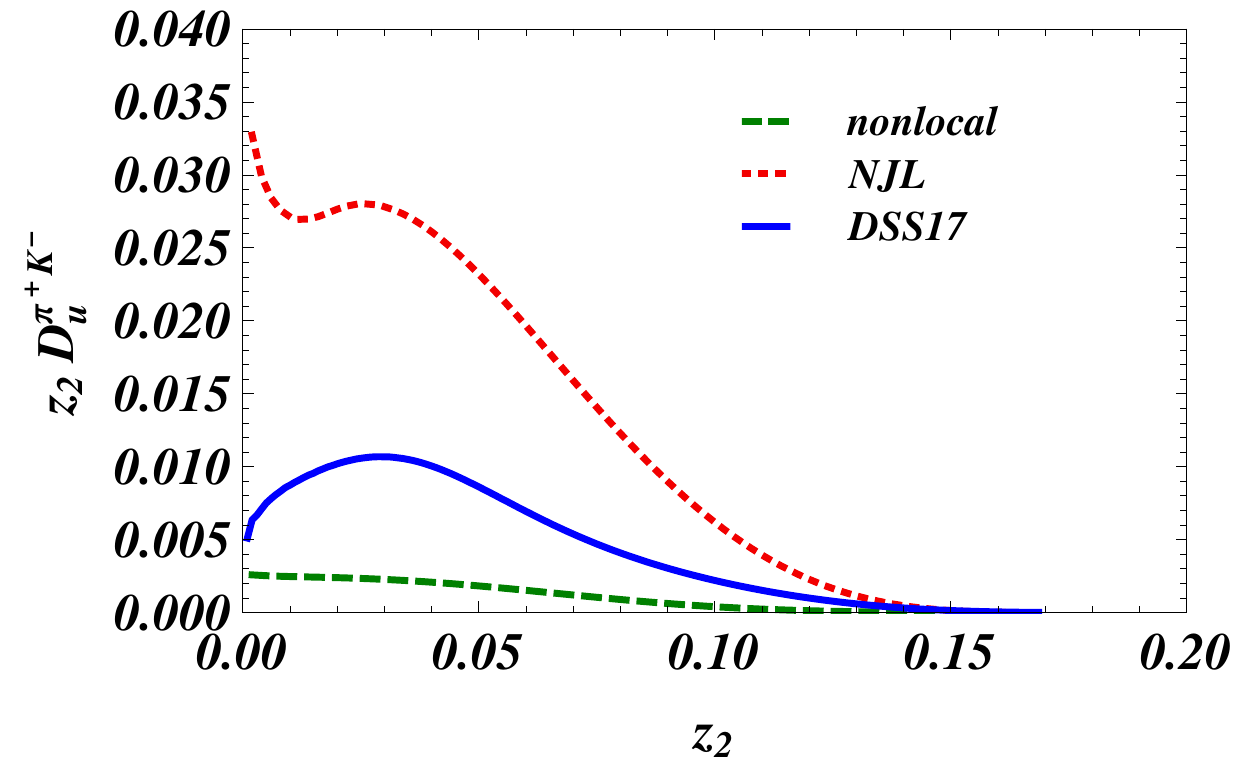}
\includegraphics[width=5.2cm]{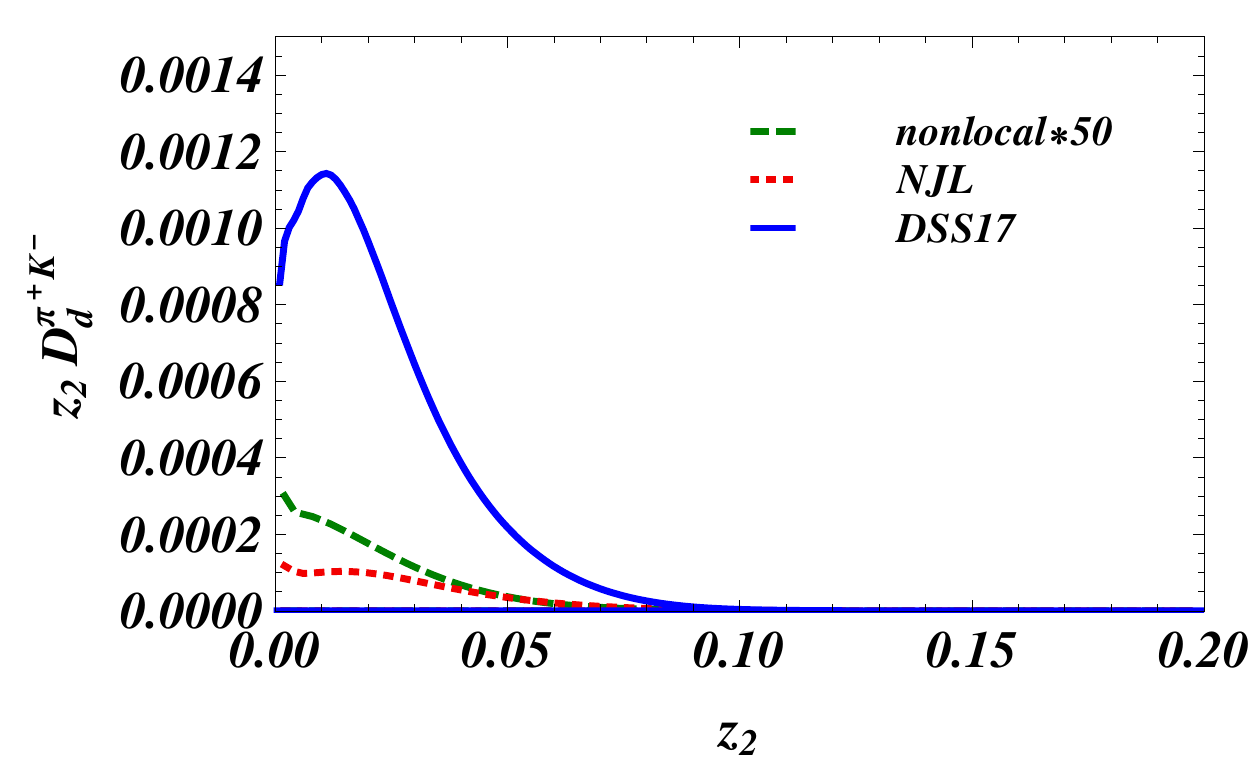}
\includegraphics[width=5.2cm]{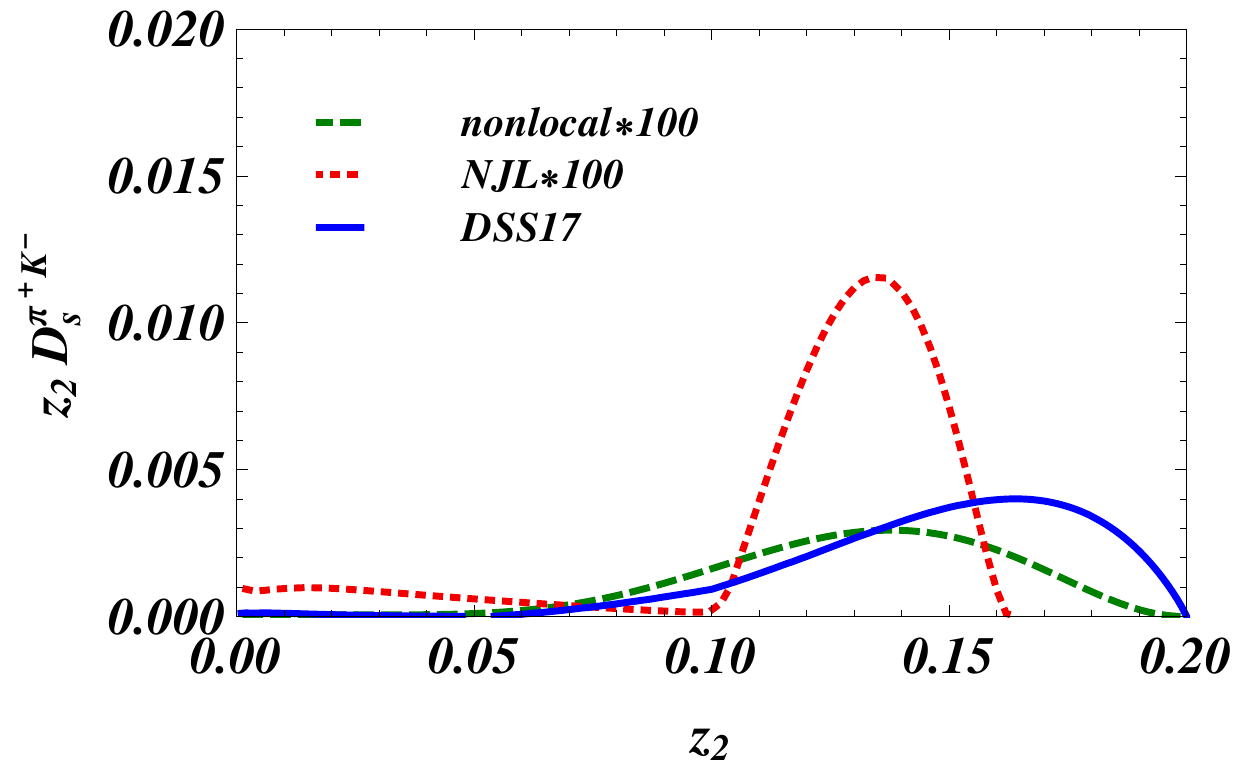}
\end{tabular}
\begin{tabular}{ccc}
\includegraphics[width=5.2cm]{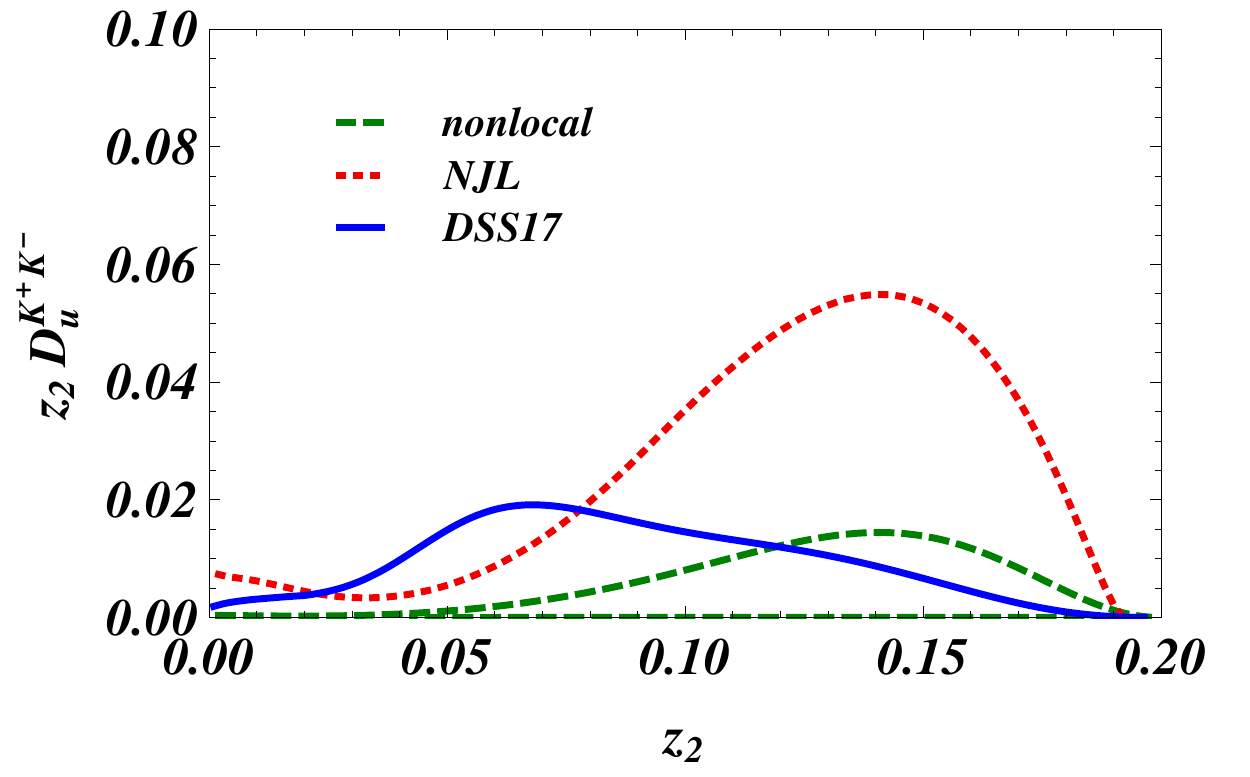}
\includegraphics[width=5.2cm]{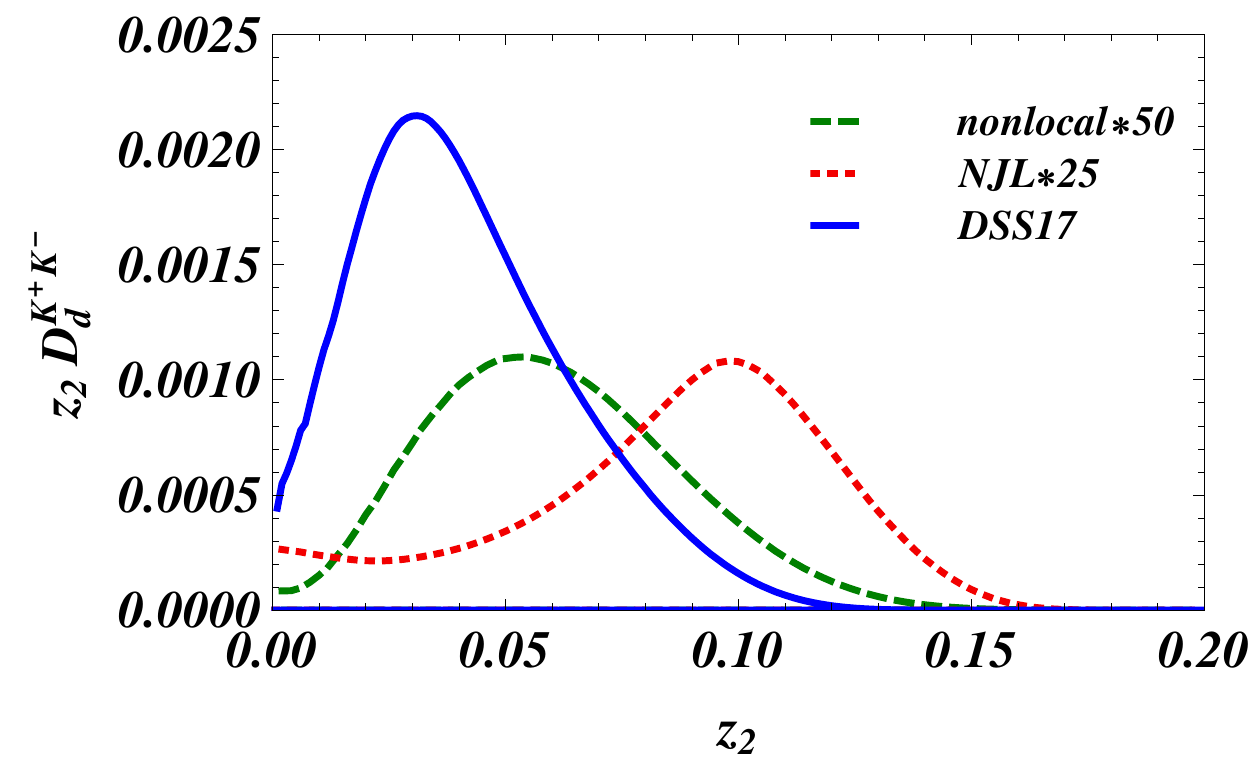}
\includegraphics[width=5.2cm]{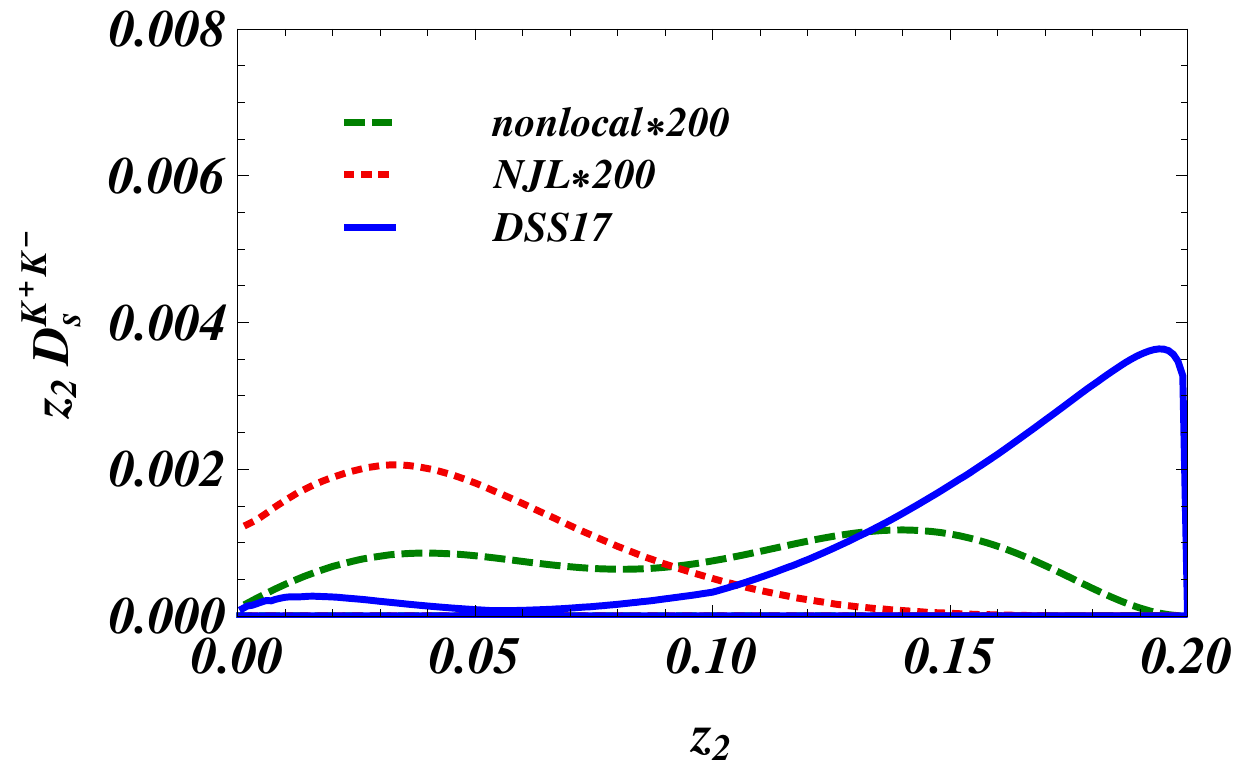}
\end{tabular}
\caption{$z_2 D^{h_1,h_2}_{q}(z_1,z_2)$ with $z_{1}=0.8$ and $Q^2=4\,\mathrm{GeV}^2$ for
(1) $(q,h_1,h_2)=(u,\pi^{+},\pi^{-})$ (left of the top row), (2) $(d,\pi^{+},\pi^{-})$
(middle of the top row), (3) $(s,\pi^{+},\pi^{-})$ (right of the top row),
(4) $(u,\pi^{+},K^{-})$ (left of the middle row), (5) $(d,\pi^{+},K^{-})$
(middle of the middle row), (6) $(s,\pi^{+},K^{-})$ (right of the middle row)
(7) $(u,K^{+},K^{-})$ (left of the bottom row), (8) $(d,K^{+},K^{-})$
(middle of the bottom row), (9) $(s,K^{+},K^{-})$ (right of the bottom row).
The dashed and solid lines denote the results of the NJL-jet model and the nonlocal chiral quark model respectively. The range of $z_2$ is from zero to 0.2.
}
\label{uDi8}
\end{figure}

We present our result of the dihadron fragmentation functions at $Q^2$=4 GeV$^2$
in Fig.~(\ref{uDi2},~\ref{uDi5},~\ref{uDi8}).
The solid lines represent the result derived from our $d^h_q$s given in Fig.~(\ref{zd}).
The dashed lines represent the result the nonlocal chiral quark model \cite{Yang:2014eca}.
The dotted lines represent the result of the NLJ model~\cite{Yang:2014eca}. These figures are differentiated from their corresponding $z_1$ values.
Fig.~(\ref{uDi2}) is the figure corresponding to $z_1=0.2$. Fig.~(\ref{uDi5})and Fig.~(\ref{uDi8}) are the figures with $z_1=0.5$ and $0.8$, respectively.
We find that our empirical results are substantially different from the results of the nonlocal chiral quark model and the NJL model. \\

We first discuss the case of $z_1=0.2$.
For the case of $u\rightarrow \pi^{+}\pi^{-}$, our empirical result is close to the result of NL$\chi$QM result except at the very low $z$ regime
where our result drops but the NL$\chi$QM does not. On the other hand, our result is much larger than the NJL model result.
Note that this channel is the most important one since its value is dominant over the other ones in magnitude.
The situation for the case of $d\rightarrow \pi^{+}\pi^{-}$ is different.
The shape of our curve is completely different from the shapes of the curves from the models.
The model results peak at high $z$ regime but ours peaks at the lower $z$ regime.
The magnitude of our result is between the results of the two models.
From observing the above two favoured uDiFFs, one finds that the non-local chiral quark model overestimates the pion multiplicities,
on the contrary, the NJL model underestimate them.
This observation agrees with the conclusion found in our previous analysis of the hadron multiplicities~\cite{Yang:2015avi}.
However, for the case of $s\rightarrow \pi^{+}\pi^{-}$, the magnitudes of our result turns out to be largest.
Although this uDiFF is unfavoured one, nevertheless its magnitude is significantly large compared with the other unfavoured ones.
It is interesting to find that the NJL model underestimate this uDiFF significantly.
Again we find the non-local chiral quark model gives a more accurate estimate than the NJL model.\\

For the cases of $u\rightarrow \pi^{+}K^{-}$ and $d\rightarrow \pi^{+}K^{-}$, we find that our results own different shapes with the two models.
Our empirical curves peak at $z_2=0.2$ but the model results decrease as $z_2$ increases.
Furthermore, our results are also much larger in magnitude. The NL$\chi$QM underestimate the two channels significantly.
The NJL provide a slightly larger result compared with the  NL$\chi$QM result,
but still much below our result. This feature is likely due to the fact that our unfavoured kaon fragmentation functions of $u$ and $d$ quarks
are overestimated as demonstrated in the previous section.
The situation is completely different when we study $s\rightarrow \pi^{+}K^{-}$. This uDiFF is favoured one and the
magnitude is quite large. We find our result peaks at lower $z$ position compared with the model results,
Moreover, the magnitude of our result is also smaller than the model results.
In particular we find the NL$\chi$QM gives the largest result.
We now turn our attention to  $u\rightarrow K^{+}K^{-}$ and $d\rightarrow K^{+}K^{-}$. We observe that our results are close to the NJL model results, on the other hand,
the NL$\chi$QM underestimate these two channels excessively. Furthermore, the NL$\chi$QM also underestimate the $s\rightarrow K^{+}K^{-}$  case, but our result here is significantly different from the NLJ model result. The peak position of the NJL model result is at $z$=0.6 but ours is at $z$=0.3.  \\

Now if we turn our attention to the $z_1$=0.5 case, the situation becomes very different.
It shows the uDiFF is very sensitive to the $z_1$ value.
For the case of $u\rightarrow \pi^{+}\pi^{-}$, the model results become larger than our result. It is found
that our result drops with increasing $z_1$ more fast than the two model results.
The shape of our curve becomes very flat compared with the case of $z_1$=0.2 and the magnitude of the result is
reduced about 85$\%$! The shapes of the model results keep approximately the same but the magnitude of the model results is reduced 50$\%$.
The situation for $d\rightarrow \pi^{+}\pi^{-}$ is very similar to the case of $u\rightarrow \pi^{+}\pi^{-}$ for our empirical result but the model results
drop surprisingly fast, they are reduced 92$\%$.
The situation for $s\rightarrow \pi^{+}\pi^{-}$ is similar to the case of $d\rightarrow \pi^{+}\pi^{-}$.
For the cases of $u\rightarrow \pi^{+}K^{-}$ and $d\rightarrow \pi^{+}K^{-}$, our empirical results drop faster than the NJL model results,
and the NL$\chi$QM results are always the smallest among three results.
For the case of $s\rightarrow \pi^{+}K^{-}$, our result becomes larger than the model results with the similar shapes of the model results,
in contrast of the case at $z_1$ =0.2 where the empirical result has different shape with the model results.
For $u\rightarrow K^{+}K^{-}$ case, our empirical result becomes smaller than the NJL model result but still much larger than the NL$\chi$QM.
Our result of $s\rightarrow K^{+}K^{-}$ is larger than the model results. Our result shows that the charged kaon pairs
come from $u$ and $s$ quark with the similar probabilities, but in the NJL model, the charged kaon pairs are mainly from $u$ quark.
Furthermore the probabilities of the production of charged kaon pair is highly suppressed in the NL$\chi$QM.\\

Finally let us study the case of $z_1$=0.8. All of the disfavoured uDiFFs become smaller than 0.01 such that we will not discuss them because their effects are negligible.
For the case of the $u$ quark DiFFs, it is obvious that the NJL model results are larger than the NL$\chi$QM result and our empirical
result. On the other hand, the uDiFFs of the $s$ quark of our approach are significantly larger than the model results, in particular, the plots of the model results in
Fig.~(\ref{uDi8}) are multiplied by the factor of 100 or 200. The situation for the $d$ quark is similar.\\

In general, we find that our empirical results are significantly different from the results of NJL model and the NL$\chi$QM in the magnitude and in the flavour dependence.
\section{Conclusion}
In this article, we obtain the elementary fragmentation functions $d^m_q(z)$ which are able to reproduce the physical single hadron fragmentation functions
$D_q^m(z)$ of the favoured channels through the single cascade algorithm. With this set of the empirical $d^m_q(z)$ we calculate the unpolarized di-hadron fragmentation
functions through the single cascade algorithm
and compare our empirical result with the results of the NJL model and the NL$\chi$QM. We find our empirical result is significantly different from the results of the models, in particular our result owns very distinct feature with the model results. Our result is believed to be more suitable to be used in the phenological study of the SIDIS and other processes with the two-hadron final states. We plan to generalize our approach to study the extended di-hadron
fragmentation functions which are crucial to extract the transversity. Another direction of the future study is to generalize to the study of the polarized fragmentation functions.

\section*{Acknowledgments}
C.-W.K was supported by the Ministry of Science and Technology of Taiwan by the grant number MOST 107-2119-M-033-002 and MOST 108-2112-M-033-004.
The work of S.i.N. was supported by the National Research
Foundation of Korea (NRF) (Grants No. 2018R1A5A1025563
and No. 2019R1A2C1005697).

\end{document}